\documentclass[11pt,a4paper]{article}

\usepackage[utf8]{inputenc}
\usepackage[T1]{fontenc}

\usepackage{amsmath}
\usepackage{color}
\usepackage{amsfonts}
\usepackage{amssymb}
\usepackage{graphicx}
\usepackage{geometry}
\usepackage{amssymb,epsfig,subfigure}
\usepackage{hyperref}
\usepackage{comment}
\usepackage[font=footnotesize]{caption}
\usepackage{dsfont}
\usepackage{braket}
\usepackage{authblk}

\numberwithin{equation}{section}
\newcommand\nn{\nonumber}
\newcommand\eea{\end{eqnarray}}
\newcommand\bea{\begin{eqnarray}}

\def\beq{\begin{equation}}
\def\eeq{\end{equation}}

\def\pa{\partial}

\newcommand{\be}{\begin{equation}}
\newcommand{\ee}{\end{equation}}
\newcommand{\ba}{\begin{align}}
\newcommand{\ea}{\end{align}}
\newcommand{\bg}{\begin{gather}}
\newcommand{\eg}{\end{gather}}
\newcommand{\bseq}{\begin{subequations}}
\newcommand{\eseq}{\end{subequations}}

\begin{document}


\bigskip
\title{\bf  A Path Integral Realization of \\ Joint $J\bar T$, $T\bar J$ and $T\bar T$ Flows
\vskip 0.5cm}

\author{ Jerem\'ias Aguilera-Damia $^a$ {\footnote{
jeremiasadlp@gmail.com }}, Victor I. Giraldo-Rivera $^b${\footnote{
vigirald@gmail.com }}, Edward A. Mazenc$^c$ {\footnote{
mazenc@stanford.edu }}, \\  Ignacio Salazar Landea $^b$ {\footnote{
peznacho@gmail.com }}, Ronak M Soni $^c$ {\footnote{
ronakms@stanford.edu }}  }

\date{
    $^a$ \small \it Centro At\'omico Bariloche and CONICET, R8402AGP, Argentina\\
    $^b$ \it Instituto de F\'isica La Plata-CONICET \& Departamento de F\'isica, Universidad Nacional de La Plata, C.C. 67, 1900, La Plata Argentina\\
    $^c$ \small \it Department of Physics, Stanford University, Stanford, CA 94305-4060, USA
    }

\maketitle

\bigskip

\begin{abstract}
We recast the joint $J\bar{T}$, $T\bar{J}$ and $T\bar{T}$ deformations as coupling the original theory to a mixture of topological gravity and gauge theory. This geometrizes the general flow triggered by irrelevant deformations built out of conserved currents and the stress-energy tensor, by means of a path integral kernel. The partition function of the deformed theory satisfies a diffusion-like flow equation similar to that found in the pure $T\bar{T}$ case. Our proposal passes two stringent tests. Firstly, we recover the classical deformed actions from the kernel, reproducing the known expressions for the free boson and fermion. Secondly, we explicitly compute the torus path integral along the flow and show it localizes to a finite-dimensional, one-loop exact integral over base space torus moduli. The dressed energy levels so obtained match exactly onto those previously reported in the literature. 
\end{abstract}
\bigskip

\pagebreak

\tableofcontents

\section{Introduction}

The Wilsonian paradigm has taught us that characterising trajectories on the space of Quantum Field Theories (QFT) is difficult but imperative to understand. These trajectories are triggered by a wide variety of deformations, whose form depends on the particular details of the conformal fixed point at which they are turned on. A UV complete scheme posses a well defined UV fixed point described by a Conformal Field Theory (CFT). In principle, we can track the flow generated by a  relevant or marginal deformation. However, this picture usually breaks down for irrelevant deformations, which give rise to non-renormalizable interactions. This obscures whether the UV physics is well defined or not.

In 2-dimensional quantum field theories, certain composite operators built out of conserved currents stand as an exception to the rule. Starting with the remarkable $T\bar{T}$ operator \cite{Zamolodchikov:2004ce} and its Lorentz preserving higher spin relatives \cite{Smirnov:2016lqw}, to the manifestly Lorentz breaking $J\bar{T}$ and $T\bar{J}$ deformations \cite{Guica:2017lia}, these special operators all share two remarkable properties. Besides being composite operators, they are unambiguously defined, {\it i.e.} free of short-distance singularities. Secondly, they give rise to exactly solvable trajectories in the space of theories.

Much recent work has succeeded in describing the dynamics $T\bar{T}$ flow and elucidating the underlying origin of its solubility. In a nutshell, general properties of the QFT stress tensor (subject to very mild assumptions) lead to the explicit factorization of its expectation value in terms of bilocal products of single trace operators.  The flow of the deformed energy levels then satisfy the inviscid Burgers equation  \cite{Smirnov:2016lqw,Cavaglia:2016oda}. \cite{Cardy:2018sdv,Cardy:2018jho} provided a rather different perspective by studying the flow infinitesimally. They showed how the $T\bar{T}$ deformation could be accounted for by coupling the seed theory to random metrics, whose action turns out to be topological. \cite{Dubovsky:2017cnj,Dubovsky:2018bmo} pushed this analysis beyond the infinitesimal regime, and recast the flow as coupling the undeformed theory to a variant of JT gravity. 

This powerful geometric interpretation further allows one to obtain the classical deformed actions in terms of a field-dependent coordinate transformation \cite{Conti:2018tca,Conti:2019dxg,Coleman:2019dvf}. For interested readers, we refer to the pedagogical review  \cite{Jiang:2019hxb} which further explains many of the concepts listed above.

The main point of this present paper is to similarly geometrize the $J\bar{T}$ and $T\bar{J}$ flows, in Euclidean space. The so-called $J\bar{T}$ and $T\bar{J}$ operators provide an equally interesting set of solvable, Lorentz breaking deformations. These operators are defined as 
\begin{align}
J \bar{T} \equiv&  \, - \epsilon_{\mu\nu}n^a J^\mu T^\nu_{\,\,a} \,, \label{jtbar}\\
T  \bar{J} \equiv& \, \epsilon_{\mu\nu}\tilde{n}^a \tilde{J}^\mu T^\nu_{\,\,a} \label{tjbar}\,,
\end{align}
where $J$ and $\tilde{J}$ are the Noether currents associated to some $U(1)$ symmetries of the underlying seed theory. $T^\mu_{\,\,a}= T^\mu_{\,\,\nu} e^\nu_a$, where $T^\mu_{\,\,\nu}$ is the stress-energy tensor and  $e^a_\mu$ the vielbein satisfying $e^{a}_{\mu}e^{b}_{\nu}\eta_{ab}=g_{\mu \nu}$.
Finally, $n,\tilde{n}$ will be taken to be two light-like vectors, satisfying
\begin{equation}
    n\cdot n= \tilde{n}\cdot\tilde{n}=0 \quad , \quad n\cdot\tilde{n}=1, \quad \delta^a_b = n^a \tilde{n}_b + \tilde{n}^a n_b
    \label{eqn:light-like-intro}
\end{equation}
These operators, in requiring two light-like vectors of opposite chirality, explicitly break Lorentz invariance.

This light-like nature in Euclidean space means the vectors cannot be real-valued. In Euclidean space the only real vector with norm $0$ is the null vector. Nonetheless, we make this choice to facilitate comparison with past papers working in Lorentzian space and choosing real light-like vectors in that context.

In terms of the above operators, the flow is described by the following expressions\footnote{The relative sign between these two equations cancels the relative sign between \eqref{jtbar},\eqref{tjbar}, and makes subsequent expressions easier to write. The $J\bar{T}+T\bar{J}$ flow, in our conventions, is $\ell_1<0,\ell_2>0$.}
\begin{equation}
\frac{\partial \log Z_{\ell}}{\partial \ell_{1}} =  \left\langle \int d^2 \sigma \, J\bar{T}\big|_{\ell_{1}} \right\rangle_{Z_{\ell}} \label{JTflow}
\end{equation} \,,
\begin{equation}
\frac{\partial \log Z_{\ell}}{\partial \ell_{2}} = - \left\langle\int d^2 \sigma \,  T\bar{J}\big|_{\ell_{2}} \right\rangle_{Z_{\ell}} \label{TJflow} \,,
\end{equation} 
with $Z_{\ell_{i}}$ denoting the partition function and $\ell_{i}$ parameters of length dimension one parametrizing the curve throughout the space of theories. Note that we are making explicit the $\ell$ dependence on both sides of \eqref{JTflow} and \eqref{TJflow},  emphasizing the recursive nature of the flow and its non-linear dependence on the deformation parameter. We will drop this excessive notation in the rest of the paper.

This flow was first introduced in \cite{Guica:2017lia}, where the deformed spectrum was obtained for a certain set of theories with vanishing $U(1)$ level. The more general case have been worked out in  
\cite{Bzowski:2018pcy}, together with an explicit construction of a holographic realization of the flow involving an $AdS$ bulk with mixed boundary conditions for the metric and Chern-Simons gauge fields.

A closely related flow, now driven by a single trace version of the $T\bar T$ operator, have been firstly proposed in \cite{Giveon:2017nie}. They ingeniously recast it as a \text{marginal} deformation of a suitable string worldsheet theory. These ideas have been further developed in \cite{Giveon:2017myj, Asrat:2017tzd, Chakraborty:2018kpr}. Generalizations of these flows involving also $J\bar T$ and $T\bar J$ operators were worked out in \cite{Chakraborty:2018vja, ApoloSong1,ApoloSong2}.   

  Moreover, the deformed energy levels for a general combination of $J\bar T$, $T\bar J$ and $T\bar T$ flows (either the single or double trace version) were recently obtained in \cite{Giveon:2019fgr, Chakraborty:2019mdf, LeFloch:2019rut, Hashimoto:2019wct}. Additionally, some generic flows have been worked out using light-cone quantization \cite{Frolov:2019xzi}. %
It has also been shown to be the unique class of flows satisfying modular covariance and some other properties in \cite{AharonyJT}, using methods similar to \cite{AharonyTT}.

The similarity to the $T\bar{T}$ deformation motivates us to seek a similar geometrical interpretation of the  $J\bar{T}$ and $T\bar{J}$ flows. We require the following of our proposal. At the classical level, it should lead to a well-defined procedure to derive the deformed classical action by means of coordinate and gauge transformations. We expect to write down a kernel as a topological action for random spacetime metrics and gauge fields. Finally, it should naturally lead to a well defined prescription for constructing the quantum partition function, encoding physically relevant properties of the deformed theory such as the energy spectrum. Our kernel does all of the above. 

~

\subsection{Summary of Results}

Here we list the main results of this paper.

\begin{itemize}

\item{
By closely following the arguments developed for $T\bar T$ deformations \cite{Dubovsky:2017cnj,Dubovsky:2018bmo}, we capture the effects of the $J\bar T$ and $T\bar J$ deformations by coupling the seed theory to topological gauge and gravitational degrees of freedom. In fact, we can easily incorporate a joint flow with $T\bar{T}$, and via a suitable limit of the parameters, even reproduce the $J\bar{J}$ deformation. 

To do so, we first write the partition function of the undeformed theory as $Z_{0}[e^a,A,\tilde{A}]$. This is a functional of the background spacetime metric (written in terms of the first order formalism vielbeins $e^{a}_{\mu}$) and background gauge fields $A_{\mu}$ and $\tilde{A}_{\mu}$ which couple to the $U(1)$ currents of the seed theory.

We obtain the partition function of the deformed theory living on a spacetime with metric vielbeins $f^{a}_{\mu}$ and background gauge fields $B_{\mu}$ and $\tilde{B}_{\mu}$ via the following path integral 

\be
Z_{\ell}[f^a, B,\tilde B]= \int \frac{De^a_\mu DY^a}{\text{vol(diff)}} \frac{DA_\mu D\alpha}{\text{vol(G)}} \frac{D\tilde A_\mu D\tilde\alpha}{\text{vol}(\tilde{\text{G}})}  e^{-S_{K} [f^a, B, \tilde B, e^a, Y^a , A, \tilde A, \alpha, \tilde\alpha]}Z_0[e^a, A, \tilde A] 
\label{defpartition}
\ee

where the kernel's action is given by 
\begin{align}
S_{K}=&\,\, \quad \frac{1}{\ell_1}\int d^2 \sigma\epsilon^{\mu\nu}\tilde{n}_a(f^a_\mu +\partial_\mu Y^a-e^a_\mu)(B_\nu+\partial_\nu \alpha- A_\nu ) \nn \\
&+\frac{1}{\ell_2}\int d^2 \sigma \epsilon^{\mu\nu}n_a (f^a_\mu+\partial_\mu Y^a-e^a_\mu)(\tilde{B}_\nu +\partial_\nu\tilde{\alpha}- \tilde{A}_\nu) \nonumber\\
&- \gamma \int d^2\sigma\, \epsilon^{\mu\nu}\left(B_\mu + \partial_\mu \alpha-A_\mu\right)\left(B_\nu + \partial_\nu \tilde\alpha-\tilde{A}_\nu\right)\,.
\label{GenJTkernel}
\end{align}
Note that $n,\tilde{n}$ have switched places; this is because of the off-diagonal decomposition of identity in \eqref{eqn:light-like-intro}.
Here, $\gamma$ is related to the deformation parameter $\lambda$ of the $T \bar{T}$ deformation as
\begin{equation}
    \lambda = \gamma \ell_1 \ell_2.
    \label{eqn:lambda}
\end{equation}

First off, readers might object to this being called a kernel, as it is not quite of the schematic form $f_{\ell}(x)=\int dy K(x,y)f_{0}(y)$. Instead, there are extra integrations over the fields $Y^a$, $\alpha$ and $\tilde{\alpha}$. The purpose of these additional integrals is to impose important constraints on the range of integration for our kernel. From our path integral perspective, these constraints lie at the heart of the solubility of these flows. 

It is helpful to introduce the distinction between the ``target'' space, where the deformed theory lives, and the ``base'' space on which the undeformed theory is defined. The base space is parametrized by the coordinates $\sigma^{\mu}$. We thus say the $f^a_\mu$ denote target space vielbeins, which will be taken as flat throughout all this work. Similarly, $B_\mu$ and $\tilde B_\mu$ denote background gauge fluxes defined in target space. The fundamental purpose of the kernel is to establish a map between base and target space variables.\footnote{By "flat", we mean $df^a=0$. We will see a proper treatment of diffeomorphism invariance in fact requires $dB=d\tilde{B}=0$ as well. Generalisations beyond the $df^a=0$ case will be considered in \cite{Mazenc:2019cfg}.}

The additional fields carry an interesting physical interpretation. Physically, the $Y^a$, $\alpha$ and $\tilde\alpha$ denote fluctuations around the above described target space backgrounds. They parametrize coordinate and gauge transformations respectively. Finally, they make the diffeomorphism and gauge invariance of the kernel manifest. 
}

\item{Section \ref{classical} first tests our proposal at the classical level, \textit{i.e.} the deformation given by replacing $ \log Z $ in \eqref{JTflow}, \eqref{TJflow} by the action $S$.  We introduce $X^{\mu}$ which play the role of a ``dynamical'' coordinate parametrizing the target space. At this stage, $B$ and $\tilde B$ do not play any role , and are set to $0$.

To probe the classical structure of the deformed theory, we can evaluate the path integral in \eqref{GenJTkernel} via saddle point. All fluctuating fields are set to their saddle point configurations, in particular the $X^\mu$. In nutshell, we rewrite the original action on a space with coordinates $X^\mu $.  In addition, we also perform a gauge transformation determined by the classical configuration of the fields $\alpha$ and $ \tilde \alpha$. This gives rise to a well-defined geometrical procedure to obtain the exact classical deformed action by solving an algebraic system of equations, in complete analogy with \cite{Coleman:2019dvf}. We successfully recover the known deformed actions for the free boson and free fermion.  Further, we propose a particular modification to account for the joint  $J\bar T$, $T\bar J$ and $T\bar T$ flow and revisit the two examples discussed above.
}

\item{We compute explicitly the path integral on the torus. The $T\bar{T}$ analog of this computation was done in \cite{Dubovsky:2018bmo}. We attempt to make this rather technical calculation as accessible as possible. We confirm the validity of our result on multiple fronts. Our deformed partition function satisfies the torus version of the flow equation:
\be
\partial_{\ell_1 }{\cal Z}= - n^a \partial_b \wedge \partial_L^a {\cal Z} \quad , \quad \partial_{\ell_2 }{\cal Z}= - \tilde n^a \partial_{\tilde b} \wedge \partial_L^a {\cal Z}
\label{diffusion}
\ee  
where $L^a_\mu$ and $b_\mu,\tilde{b}_\mu$  denote respectively the lengths of the torus and the holonomies of the background gauge fields $B,\tilde{B}$ that couple to $J,\tilde{J}$.
A similar equation was worked out using modular properties in \cite{AharonyJT}.

We have also defined ${\cal Z}= Z/{\cal A}$ with ${\cal A}$ the corresponding area of the torus.

}

\item{ The explicit evaluation of \eqref{defpartition} for the torus provides a further stringent test because it allows us to access the spectrum of the deformed theory. Indeed, by setting the target space gauge fields to zero, we know the torus partition function can be written as the sum
\be
Z=\sum_{n} e^{2\pi i \tau_1 R P_n -2\pi \tau_2 R E_n}\,,
\label{gentorus}
\ee 
with $R$ denoting the length of the spatial circle of the target space torus. $\tau_1,\tau_2$ are its modular parameters and $P_{n} =k_n/R$ the quantized spatial momentum ({\it i.e.} $k_n\in {\mathbb Z}$). From \eqref{gentorus}, we can extract  the deformed energy levels given by 
\begin{align}
E_{n}= &\frac{1 }{2\pi^2k(\ell_1+\ell_2)^2}\Big(R-\ell_1 \pi Q_n +\ell_2\pi\tilde{Q}_n+ 2\pi^2 k (\ell_1+\ell_2)(\ell_1-\ell_2)\frac{k_n}{R}\, + \label{doubledef}\\
&\left.\sqrt{(R-\ell_1 \pi Q_n+\ell_2 \pi \tilde{Q}_n)^2-8\pi^2 k (\ell_1+\ell_2)^2\left(\varepsilon_{0,n}^{(R)}+k_n\ell_2\frac{R-\ell_1\pi(Q_n+\tilde Q_n)}{R(\ell_1+\ell_2)}\right)}\right)\nn
\end{align}
where the $\varepsilon^{(R)}_{0,n}$ are the (dimensionless) right-moving energies of the seed theory and $Q_n$($\tilde Q_n$) the charge associated to $J$($\tilde J$). \footnote{We have tried to adhere as closely as possible to conventional notation. There are unfortunately too many $R$'s in this business: the one in the superscript denotes "right moving". Otherwise, it is the length of the spatial cycle of the torus.}  Additionally, $k$ denotes the level of the chiral algebra satisfied by the currents. Up to notational conventions, this agrees with the spectrum in the existing literature, obtained via very different approaches. Moreover, we can similarly compute the spectrum along the joint flow of $J\bar T$ and $T\bar T$, again finding precise agreement with previous results. 
}

\item{  Not only does the torus path integral localize to a finite-dimensional one, it is furthermore one-loop exact.  This means the exact integral can be computed from its saddle point approximation, even though it is not Gaussian. From the path integral point of view, this provides a further understanding for the solubility of the $J\bar{T}$ and $T\bar J$ flow. Our kernel shares this remarkable property with its $T\bar{T}$ predecessor \cite{Dubovsky:2018bmo} . More importantly, we extract from the saddle-point equation important physical intuition and make contact with previous definitions of "chiral" charge appearing in \cite{Bzowski:2018pcy,Chakraborty:2018vja}.
}

\end{itemize}

\subsubsection*{Note Added}

Similar aspects of these flows are also considered in \cite{Anous:2019osb}. We thank the authors for kind correspondence on this matter.

\section{An Introduction to the Kernel} \label{kernel}
The main proposal of this paper is that a joint flow under $T \bar{T}$, $T \bar{J}$ and $J \bar{T}$ is implemented by the path integral
\begin{align}
    Z_{\ell_1,\ell_2,\gamma} [f,B,\tilde{B}] &= \int \frac{De^a_\mu DY^a}{\text{vol(diff)}} \frac{DA_\mu D\alpha}{\text{vol(G)}} \frac{D\tilde A_\mu D\tilde\alpha}{\text{vol}(\tilde{\text{G}})}  e^{-S_K [f^a, B, \tilde B, e^a, Y^a , A, \tilde A, \alpha, \tilde\alpha]}Z_0[e^a, A, \tilde A] \label{full-theory} \intertext{where}
    S_K =& \,\, \quad \frac{1}{\ell_1}\int d^2 \sigma\epsilon^{\mu\nu}\tilde{n}_a(f^a_\mu +\partial_\mu Y^a-e^a_\mu)(B_\nu+\partial_\nu \alpha- A_\nu ) \nn \\
    &+\frac{1}{\ell_2}\int d^2 \sigma \epsilon^{\mu\nu}n_a (f^a_\mu+\partial_\mu Y^a-e^a_\mu)(\tilde{B}_\nu +\partial_\nu\tilde{\alpha}- \tilde{A}_\nu) \nonumber\\
    &- \gamma \int d^2\sigma\, \epsilon^{\mu\nu}\left(B_\mu + \partial_\mu \alpha-A_\mu\right)\left(B_\nu + \partial_\nu \tilde\alpha-\tilde{A}_\nu\right)\,.\nonumber \intertext{and}
    &\partial_{[\mu} f_{\nu]}^a = \partial_{[\mu} B_{\nu]} = \partial_{[\mu} \tilde{B}_{\nu]} = 0. \nonumber
\end{align}
The purpose of this section is to introduce some important properties of this kernel. Along the way, we will carefully define the various quantities appearing in it.

Before delving into the full story, we need to list a few choices that complete the definition of this theory.
\begin{enumerate}
    \item All fields that appear in \eqref{full-theory} are `single-valued.' On the torus, this actually means single-valued. On non-compact spaces, this means that it vanishes at the boundary. Either way, integration by parts in terms of these objects never generates boundary terms.
    \item We impose the interpretation $Y^a = f_\mu^a \xi^\mu$, where $\xi^\mu$ is a single-valued vector field; the motivation for this is elucidated later in this section, and this interpretation will be crucial in doing the path integral.
    
    \item On a related note, we require $\alpha,\tilde{\alpha}$ to be valued in the respective gauge groups. The relation between these requirements on $\alpha,\tilde{\alpha}$ and $Y$ should become clear as this section progresses.
    \item On the torus, the gauge group volumes are only volumes of the groups of gauge transformations continuously connected to the identity. This is intimately linked with the single-valuedness of the fields, in a way that will be explained in section \ref{quantum}.
\end{enumerate}
We can now move on to discussing more interesting qualitative questions.

The first important fact to notice about this theory is key to its topological nature and underlies the factorisation of the bi-linear operators that define the flows. The fields $Y^a,\alpha,\tilde{\alpha}$ act as Lagrange multipliers setting the curvatures of the fixed and dynamical gauge fields to coincide:
\begin{equation}
    \partial_{[\mu} (f-e)_{\nu]}^a = \partial_{[\mu} (B-A)_{\nu]} = \partial_{[\mu} (\tilde{B}-\tilde{A})_{\nu]} = 0.
    \label{localisation}
\end{equation}
Together with the conditions that have been imposed on the external fields $f,B,\tilde{B}$, this implies the path integral runs only over flat vielbeins and gauge fields.
The full path integral \eqref{full-theory} therefore reduces to integral only over global modes of the vielbeins and gauge fields.
We can parametrize the global modes of the vielbeins as overall scale, a uniform tangent-space rotation, and the torus moduli \cite{Dubovsky:2018bmo}.  The holonomies around the non-trivial cylces of the torus play the analogous role for the gauge fields.

There are two useful pictures to keep in mind when parsing the meaning of this path integral.
In the first, we simply view the kernel as arising from some complicated action with sources $f,B,\tilde{B}$ for the currents $T,J,\tilde{J}$. Because these are conserved currents, these background sources have special names: the vielbeins of the manifold and the background gauge fields, respectively.
This is an unreasonably hard-nosed view. Of course, whatever words we associate with the kernel, they must be compatible with this picture. 

A more intuitive picture arises by noticing that one of the dynamical variables is yet another set of vielbeins $e^a$. These can be thought of as defining the metric on a secondary manifold on which the seed theory lives.
We call the original, non-dynamical, manifold on which the deformed theory lives the \emph{target space} and this new ``dynamical'' manifold on which the seed theory lives the \emph{base space}.
Nice as this picture is, it raises an important question.  If there are two manifolds, there should exist two sets of diffeomorphisms, and similarly two sets of $U(1)$ transformations. 

These two sets of transformations have different interpretations.
Invariance under $U(1)$ transformations and diffeomorphisms of the non-dynamical target space fields encodes the conservation of the deformed currents. This is nothing new: invariance under such background field transformations always has this interpretation in QFT.
As for the base space, invariance under $U(1)$ transformations of the fields living here is a genuine gauge-invariance. The conservation of the seed currents is a necessary condition for this $U(1)$ symmetry to be gauged. The same holds for base space diffeomorphism-invariance. 

Let us see how the kernel encodes these transformations. We begin first with the $U(1)$'s. 
The action in \eqref{GenJTkernel} is readily seen to be invariant under a base space $U(1)$ gauge transformation of the form
\begin{align}
  \delta_{g} A_\mu &= \partial_\mu g \nonumber\\
  \delta_{g} \alpha &= g \nonumber\\
  \delta_{g} Z_{0} &= 0,
  \label{eqn:u1-g-tr-int}
\end{align}
and a target space transformation of the form
\begin{align}
  \delta B &= d g_{TS} \nonumber\\
  \delta \alpha &= - g_{TS}.
  \label{eqn:u1-g-tr-ts-int}
\end{align}
The exact same transformation rules, with tilded quantities instead, describes the other $U(1)$. 
$\alpha,\tilde{\alpha} $ play the role of compensator fields. They transform linearly under these symmetries. They parametrize the difference in $U(1)$ frames between the two manifolds.  More plainly, these transformations suggest that moving any charged field from the base to the target space requires a gauge transformation with the parameter $\alpha, \tilde{\alpha}$. This underpins the construction of classical actions in section \ref{classical}.

We now turn to diffeomorphisms. These are more subtle. The action is manifestly invariant under base space diffeomorphisms, if \text{all} the fields transform under the usual rules compatible with a reparametrization of the coordinates $\sigma^{\mu}$. This includes a transformation of the background sources $f^a \rightarrow f'^a , B \rightarrow B'$ and $\tilde{B} \rightarrow \tilde{B'}$. On the face of it, this would naively give $Z[f'^a, B' , \tilde{B}']$. What we need to show is that $Z[f'^a, B' \tilde{B}']= Z[f^a,B,\tilde{B}]$. This is tantamount to target space diffeomorphism invariance, whose infinitesimal version is just the conservation of the deformed stress tensor.

Under these target space diffeomorphisms, our fields transform as 
\begin{align}
  \delta f_\mu^{a} &= (\mathcal{L}_{\xi} f^{a})_\mu = \xi^\nu \partial_\nu f_\mu^a + (\partial_\nu \xi^\nu) f_\mu^a \nonumber\\
  \delta Y^{a} &= - \xi^\mu f_\mu^{a} \nonumber\\
  \delta B_\mu &= (\mathcal{L}_{\xi} B)_\mu.
  \label{eqn:ts-diff-trs-int}
\end{align}

However, invariance still does not follow straightforwardly.
To show this, we focus only on the term proportional to $\ell_1$ ---  the other terms are similar and follow the same argument. Resorting to form language so as to avoid a proliferation of indices, we find that this part of the action transforms as
\begin{align}
  \delta_{\xi} S_{\ell_{1}} &= \frac{1}{\ell_{1}} \int n_{a} i_{\xi} df^{a} \wedge (B + d\alpha - A) + n_{a} (f + dY -e)^{a} \wedge ( i_{\xi} dB + d i_{\xi}B) \nonumber\\
  &= \frac{1}{\ell_{1}} \int n_{a} i_{\xi} df^{a} \wedge (B + d\alpha - A) + n_{a} i_{\xi} B (df-de)^{a} + n_{a} (f+dY-e)^{a} \wedge i_{\xi} dB \nonumber\\
  &\xrightarrow{df^{a} = dB = 0} 0.
  \label{eqn:ts-diff-ds-ints}
\end{align}
Here, we have used the fact that $Y^a$ imposes $\partial_{[\mu} (f-e)_{\nu]}^a = 0.$\footnote{We have also integrated by parts. Single-valuedness of the fields, along with $df^a=dB=d\tilde{B}=0$, allows us to discard the boundary terms that arise.}
What we have found is that \eqref{eqn:ts-diff-trs-int} is a symmetry if and only if
\begin{equation}
    \partial_{[\mu} f_{\nu]}^a = \partial_{[\mu} B_{\nu]} = \partial_{[\mu} \tilde{B}_{\nu]} = 0.
    \label{eqn:kernel-reqts}
\end{equation}
In other words, the target space manifold and background gauge fields must be flat. This restriction reflects the kernel we are using necessitates additional terms to accommodate non-zero curvature of the frame-field and gauge connections.\footnote{The generalisation, in the pure $T \bar{T}$ case, to curved target spaces will be discussed in \cite{Mazenc:2019cfg}. The generalisation to non-flat target space gauge fields is currently being explored.}

Restricting our attention to the case where \eqref{eqn:kernel-reqts} hold, \eqref{eqn:ts-diff-ds-ints} shows the $Y$'s act as compensator fields for target space diffeomorphisms. Paralleling the $U(1)$ discussion, we conclude the $Y$s parametrize the difference in space-time frames. We may therefore define target space coordinates as
\begin{equation}
    X^\mu = \sigma^\mu + f^\mu_a Y^a.
    \label{eqn:ts-coords}
\end{equation}

In these coordinates, the target space vielbein is
\begin{equation}
    (X^*f)_\mu^a = \partial_\mu X^a = f_\mu^a + \partial_\mu Y^a, \quad X^a \equiv f_\mu^a X^\mu.
    \label{eqn:ts-vielbein-X}
\end{equation}
This follows the notation used in \cite{Dubovsky:2018bmo}. 
Our interpretation of the $Y^a$'s as compensator fields for diffeomorphism originally provided the backbone for deriving deformed classical actions in \cite{Coleman:2019dvf}. Further, the interpretation \eqref{eqn:ts-coords} is actually necessary for a proper definition of the path integral. It will inform our treatment of an important zero-mode in section \ref{quantum}.

\subsection{An Intuitive Derivation of the Kernel from an Infinitesimal Analysis}
Having discussed at length the various symmetries and redundancies of the kernel, we provide further intuition via a derivation along the lines of
\cite{Cardy:2018sdv}. 
We will only explicitly write equations for the $\ell_1$ flow. All other terms behave similarly. The reader should envision all the below manipulations are happening for all three sectors at once.

The analysis begins by solving the flow equation \eqref{JTflow} to first order in $\delta \ell_1$.
We use the fact the currents can be thought of as the response to a change in the corresponding background field,\footnote{$\det f = \sqrt{g}$; since we shall be concerned with flat manifolds with Cartesian coordinates throughout, we will drop this factor subsequently.}
\begin{align}
    \langle T^\mu_a (\sigma) \rangle &= \frac{1}{\det f} \frac{\delta }{\delta f_\mu^a (\sigma)} \log Z[f,B] \nonumber\\
    \langle J^\mu (\sigma) \rangle &= \frac{1}{\det f} \frac{\delta}{\delta B_{\mu} (\sigma)} \log Z[f,B]
\end{align}
to write the formal expression
\begin{equation}
    Z_{\ell_1 + \delta \ell_1} [f,B] = e^{- \delta \ell_1 \int d^2 \sigma \epsilon_{\mu\nu} n^a \frac{\delta }{\delta B_{\mu} (\sigma) } \frac{\delta }{\delta f_{\nu}^a (\sigma)}  } Z_{\ell_1} [f,B].
    \label{eqn:inf-def-Z-1}
\end{equation}
This expression, while intuitively appealing, is not entirely well-defined, because of the two coincident functional derivatives.

 \cite{Cardy:2018sdv} therefore suggested performing a Hubbard-Stratonovich transformation leading instead to 
\begin{equation}
    Z_{\ell_1 + \delta \ell_1} [f,B] = \int D \delta e D \delta A\ e^{- \frac{1}{\delta \ell_1} \int d^2 \sigma \epsilon^{\mu\nu} n_a \delta e^a_{\mu} (\sigma)   \delta A_{\nu}(\sigma)} Z_{\ell_1} [f-\delta e, B - \delta A].
    \label{eqn:inf-def-Z-HS}
\end{equation}
The crucial step here was to absorb the linear terms generated by the Hubbard-Stratonovich into a change in the partition function.
An infinitesimal step along the flow translates to an integral over small fluctuations of the vielbeins and gauge fields relating base and target space variables. Comparing with our kernel, we can identify $f= e_{base} + \delta e$ an $B = A_{base} + \delta A $.

There is as yet no hint of the compensator fields $Y^a$, $\alpha$ and $\tilde{\alpha}$.  Conversely, there is no division by the volume of gauge transformations and diffeomorphisms.
We can incorporate them into this analysis by noting the integral \eqref{eqn:inf-def-Z-HS} is Gaussian. Its value is therefore entirely controlled by the saddle point,
\begin{align}
    \delta A_{*\mu} &= \delta \ell_1 \epsilon_{\mu\nu} n^a T^\nu_a \nonumber\\
    n_a \delta e_{*\mu}^a &= -\delta \ell_1 \epsilon_{\mu\nu} n^a J^\nu.
    \label{eqn:cardy-saddle-pt}
\end{align}
These saddle points have the special property that
\begin{equation}
    \partial_{[\mu} \delta A_{* \nu]} = \partial_{[\mu} \delta e^a_{* \nu]} = 0.
    \label{eqn:cardy-saddle-pt-flatness}
\end{equation}
Because the integral is Gaussian and therefore completely controlled by the saddle point, we may as well restrict our integration to variations satisfying \eqref{eqn:cardy-saddle-pt-flatness}. We add Lagrange multipliers imposing them as constraints.  Being linear in the fields, these constraints do not affect the fluctuation determinant around the saddle either. 
This leads to
\begin{equation}
    Z_{\ell_1 + \delta \ell_1} [f,B] = \int D\delta e D Y\ D \delta A D\alpha\ e^{- \frac{1}{\delta \ell_1} \int \epsilon^{\mu\nu} n_a (\delta e_\mu^a \delta A_\nu - \alpha \partial_{\mu} \delta e^a_{\nu} - Y^a \partial_{\mu} \delta A_\nu) } Z_{\ell_1} [f-\delta e, B-\delta A],
\end{equation}
which is clearly the infinitesimal version of \eqref{GenJTkernel}. To be clear, some choices have been made here in the normalisation of the Lagrange multipliers.

Finally, we note this object now exhibits gauge-invariance under both diffeomorphisms and $U(1)$ transformations. Therefore, we should divide by the volume of those groups. 

\section{Deformed Classical Action from the Kernel}\label{classical}

In this section, we use the classical limit of the kernel \eqref{GenJTkernel} to derive the classical action  of the deformed theory on $\mathbb{R}^2$.
This derivation follows from the interpretation, motivated in section \ref{kernel}, that the $\alpha,\tilde{\alpha},Y$ fields parametrize the difference between the coordinate systems and $U(1)$ frames of the target and base spaces.
We use the equations of motion derived from the action. This gives us the saddle-point values of these fields. We then perform the corresponding $U(1)$ and coordinate transformations on the fields of the seed theory, and evaluate the full action in terms of these transformed fields.

We begin with a review of the procedure in the case of the $T \bar{T}$ deformation, and then move on to our new case of interest.

\subsection{Review of the $T\bar{T}$ case}

Since this section's line of reasoning closely parallels that used for the $T\bar{T}$ deformation, we briefly review the relevant arguments. We will be very schematic; a detailed explanation can be found in \cite{Coleman:2019dvf}, but the idea originates in \cite{Conti:2018tca,Conti:2019dxg}.

The path integral version of a $T \bar{T}$-deformed theory \cite{Dubovsky:2017cnj,Dubovsky:2018bmo} closely resembles our path integral. The difference lies primarily in the fact their integral runs only over the vielbeins $e$ and the diffeomorphism compensators $Y^a$, with action
\be
S_{K}^{T\bar{T}} = -\frac{1}{2\lambda} \int d^2\sigma \epsilon^{\mu\nu}\epsilon_{ab} (\partial_\mu X^a-e^a_\mu)(\partial_\nu X^b-e^b_\nu) \,,
\label{TTkernel}
\ee
Once evaluated on-shell, this takes the form of the $T\bar{T}$ operator.\footnote{This kernel action needs of course be supplemented by the original action of the seed $S_0$, when deriving the equations of motion for $e$.}  $\lambda$ is a dimensionful coupling with dimensions of $({\rm length})^2$. It parametrizes the $T\bar{T}$ flow. Integrating by parts, it is easy to check the fields $X^a$ ensure the flatness condition $\epsilon^{\mu\nu}\partial_\mu e^a_\nu =0$, or in form language $de^a=0$.

The saddle-point equations that follow from the action \eqref{TTkernel} read
\begin{equation}
    \partial_\mu X^a = e_\mu^a + \lambda \epsilon_{\mu\nu} \epsilon^{ab} T_b^\nu [\phi(\sigma)],
    \label{eqn:tt-eom}
\end{equation}
with $\phi(\sigma)$ the fields of the seed theory.
These equations are linear in $X$. In principle, this is the solution for $X$ in terms of $\phi(\sigma)$. This is however not the form we are after. 
Since we want to perform coordinate transformations on the fields,
\begin{equation}
    \phi (\sigma) = X^* \tilde{\phi} (X),
    \label{eqn:coord-pull}
\end{equation}
where $X^*$ is shorthand for any tensor transformations the field must undergo,\footnote{Despite this notation being inspired by a pullback, we emphasise that this expression is \emph{not} a pullback.} it is rather cumbersome to have the coordinate transformation written in terms of the base space field $\phi$.
Instead, we rewrite the base space stress tensor in terms of the \emph{target space field} $\tilde{\phi}(X)$ to find
\begin{equation}
    \partial_\mu X^a = e_\mu^a + \lambda \epsilon_{\mu\nu} \epsilon^{ab} T_b^\nu [X^* \tilde{\phi}(X)],
    \label{eqn:tt-eom-2}
\end{equation}
where $X^* \tilde{\phi} (X)$ is the base space field written in terms of the target space fields.
 In principle, the LHS involves $X^a$ while the RHS involves $X^\mu$. These are related as in \eqref{eqn:ts-vielbein-X}. We take the special case of $\mathbb{R}^2$, where the vielbein relating them is
\begin{equation}
    f_\mu^a = \delta_\mu^a,
    \label{eqn:ts-vielbein-r2}
\end{equation}
and so they effectively coincide.

We can write the deformed classical action as the sum of the original action and the kernel action, reformulated as living on the target space 
\begin{equation}
    S_\lambda [\tilde{\phi}(X)] = \int \frac{d^2 X}{\det (\partial_{\mu} X)} \left\{ \mathcal{L}_0 [e, X^* \tilde{\phi} (X)] + \mathcal{L}_K^{T \bar{T}} [e, \partial_\sigma X \cdot \partial_X\{X^* \tilde{\phi} (X)\}, X^* \tilde{\phi} (X)] \right\}.
\end{equation}
\cite{Conti:2018tca,Coleman:2019dvf} only considered the cases where $\phi(\sigma) = \tilde{\phi} (X)$, so that the ``$X^*$'' acted as identity. We will similarly restrict to these cases.

\subsection{Generalization to $J\bar{T}$, $T\bar{J}$ and $T \bar{T}$ Deformations} \label{JTbar+TJbar}

Generalizing the above arguments to our case of interest is surprisingly straightforward. Consider a CFT whose action $S_0$ gives rise to two (or at least one) $U(1)$ symmetries. Denote by $J^\mu$ and $\tilde{J}^\mu$ the associated Noether currents. Following the steps reviewed above for the $T\bar{T}$ case, we gauge both spacetime and $U(1)$ symmetries by coupling them to ``dynamical'' vielbein $e^a_\mu$ and gauge fields $A_\mu$ and $\tilde{A}_\mu$. This promotes the action
\be
S_0[\phi] \to S_0[\phi,e^a_\mu,A_\nu,\tilde{A}_\mu] \,,
\label{minimal}
\ee 
where $\phi$ collectively denotes the original matter fields in the seed CFT.

For $J\bar{T}$ and $T\bar{J}$, the kernel action for $B=\tilde{B}=0$ and similarly $f_\mu^a = \delta_\mu^a$ is
\begin{align}
S_{K}=& \frac{1}{\ell_1}\int d^2 \sigma\epsilon^{\mu\nu}\tilde{n}_a(\partial_\mu X^a-e^a_\mu)(\partial_\nu \alpha- A_\nu ) +\frac{1}{\ell_2}\int d^2 \sigma \epsilon^{\mu\nu}n_a (\partial_\mu X^a-e^a_\mu)(\partial_\nu\tilde{\alpha}- \tilde{A}_\nu) \,. \nonumber\\
&\quad - \gamma \int \epsilon^{\mu\nu} (\partial_\mu \alpha - A_\mu) (\partial_\nu \tilde{\alpha} - \tilde{A}_\nu).
\label{JTkernel}
\end{align}
We take $n$, $\tilde{n}$ to be normalized ``light-like'' vectors, that is
\be
n\cdot n= \tilde{n}\cdot\tilde{n}=0 \quad , \quad n\cdot\tilde{n}=1 \,.
\ee 
$\sigma^\mu$ again denote the base space coordinates. The total action for all the fields thus becomes
\be
S= S_K[X^a,e^a_\mu, A_\mu,\tilde{A}_\mu,\alpha,\tilde\alpha] + S_0[\phi, e^a_\mu,A_\mu,\tilde{A}_\mu] \,. \label{totalaction}
\ee
We have already noted that $\alpha,\tilde \alpha $ play the role of Lagrange multipliers imposing the constraints  $\epsilon^{\mu\nu}\partial_{\mu}e^a_\nu=0$. 
\eqref{JTkernel} shows the combinations $\tilde{n}\cdot X$ and $n\cdot X$ also act as Lagrange multipliers. They, in turn, enforce the vanishing field strengths for $A_\mu$ and $\tilde{A}_\mu$, respectively.

The equations of motion obtained by varying \eqref{totalaction} w.r.t. the gauge fields and the vielbein are
 \begin{align}
\partial_\mu\alpha-A_\mu &=\ell_1 \, \epsilon_{\mu\nu}n^a T^\nu_a \nn\\
\partial_\mu\tilde\alpha-\tilde{A}_\mu &= \ell_2 \, \epsilon_{\mu\nu}\tilde{n}^a T^\nu_a \nn\\
\tilde{n}_a (\partial_\mu X^a-e^a_\mu) &=-\ell_1 \,  \epsilon_{\mu\nu} J^\nu - \gamma \ell_1 \ell_2 \epsilon_{\mu\nu} \tilde{n}^a T^\nu_a \nn\\
n_a (\partial_\mu X^a-e^a_\mu) &=-\ell_2 \, \epsilon_{\mu\nu} \tilde{J}^\nu + \gamma \ell_1 \ell_2 \epsilon_{\mu\nu} n^a T^\nu_a. \label{EOM}
\end{align}

We wish to obtain the classical Lagrangian for the theory defined on the plane.\footnote{The subsequent gauge choice would need to be modified for spaces with non-trivial cycles, where we cannot globally choose the connection to vanish.} In order to solve \eqref{EOM}, we choose a gauge such that  
\begin{align}
 e^a_\mu &= \delta^a_\mu \label{gaugefixing} \,,\\
 A_\mu=&\tilde{A}_\mu=0 \,,\nn 
\end{align}
consistent with the constraints imposed by the $X^a$, $\alpha$ and $\tilde{\alpha}$ fields.

Now, to find the deformed action, we need to solve the system of equations \eqref{EOM}.
As in the $T \bar{T}$ case, they look entirely linear in $X,\alpha,\tilde{\alpha}$, as long as the fields are thought of as living on the base space coordinates and base space $U(1)$ frame.
However, we are interested in the solutions in terms of fields living on the target space coordinates and $U(1)$ frame,
\begin{equation}
    \phi(\sigma) = X^* \tilde{\phi}(X)|_\alpha.
    \label{eqn:field-pull}
\end{equation}
Here, the right hand side is the base space field written in terms of target space coordinates $X$ and $U(1)$ transformed by the amounts $\alpha,\tilde{\alpha}$.
In terms of this $\tilde{\phi}$, the stress tensor and the currents depend non-trivially on $X,\alpha,\tilde{\alpha}$, and the equation is suitably non-linear.
As the transformed quantities depend on $\alpha$ and $\tilde\alpha$, \eqref{EOM} is a system of 8 equations for 8 variables. The 8 variables are the $2\times 2$ "gauge matrices" ($\partial_{\sigma^{\mu}}\alpha$ , $\partial_{\sigma^{\mu}}\tilde\alpha$) and the Jacobian $\partial_{\sigma^{\mu}} X^a$. These can be solved for algebraically.

~

To derive the classical deformed action in our case, we need simply to evaluate the action \eqref{totalaction} on shell in terms of the gauge transformed fields living on the target space coordinates $X$:
\begin{align}
S_{def} &= \int \frac{d^2 X}{\det(\partial_\sigma X)} \left. \left( S_0 -\ell_1 \epsilon_{\mu\nu}n^a J^\mu T^\nu_a -\ell_2 \epsilon_{\mu\nu}\tilde{n}^a \tilde{J}^\mu T^\nu_a  -  \gamma \ell_1 \ell_2 \left[ \epsilon^{ab} \epsilon_{\mu\nu} T_a^\mu T_b^\nu \right]\right)\right|_\alpha[\sigma(X)]  \,.
\label{defaction}
\end{align}

To see this all in action, we apply our formalism to several concrete examples in the following section.

\subsection{Some Examples of $J \bar{T}$ + $\bar{J} T$ Deformations}\label{sec:examples}
In this section, we show how the steps outlined in the previous section work for the special case of $\gamma=0$.
We will work in complex coordinates on both manifolds, for the both space-time and target/tangent space\footnote{Recall the discussion around \eqref{eqn:ts-vielbein-r2}, which explains why these two a priori different spaces are effectively identified because of the simplicity of $f$.} coordinates.
Complex target space coordinates will be denoted by $z^\mu \in (z,\bar{z})$ whereas we use complex coordinates $w^{\mu} \in (w,\bar{w})$ for the base space.
The coordinates are normalised so that
\begin{equation}
    g_{z \bar{z}} = g_{w \bar{w}} = \frac{1}{2}.
    \label{complex-metric}
\end{equation}

The flat space vielbein is gauge-fixed to be diagonal, {\it i.e.} its non-zero components are $e^z_w=e^{\bar{z}}_{\bar{w}}=1$. In this coordinates, the ``lightlike'' vectors $n$ and $\tilde{n}$ are such that
\be
n_z=\tilde{n}_{\bar{z}}= \frac{1}{\sqrt{2}}\quad , \quad n_{\bar{z}}=\tilde{n}_z=0 \,.
\ee
Raising the index, we find
\begin{equation}
    n^z = \tilde{n}^{\bar{z}} = 0, \quad n^{\bar{z}} = \tilde{n}^z = \sqrt{2}.
\end{equation}
 
Equations \eqref{EOM} now specify the 2$\times$2 Jacobian matrix with elements $\frac{\partial z^a}{\partial w^\mu}$ and the "gauge matrices" ($\partial_{w^{\mu}}\alpha$ , $\partial_{w^{\mu}}\tilde\alpha$) in terms of the currents and the stress-energy tensor defined w.r.t. the base space variables $(w,\bar{w})$. More specifically,
\begin{align}
\frac{\partial z}{\partial w} = 1-2\sqrt{2} \ell_2 \tilde{J}_{w}^g \quad &, \quad
\frac{\partial z}{\partial \bar{w}} = 2\sqrt{2} \ell_2 \tilde{J}_{\bar{w}}^g\nn\\
\frac{\partial \bar{z}}{\partial w} = -2\sqrt{2}\ell_1 J_{w}^g \quad &, \quad
\frac{\partial \bar{z}}{\partial \bar{w}} =1 +2\sqrt{2}\ell_1 J_{\bar{w}}^g\label{complexdiff}\\
\partial_w\alpha =2\sqrt{2} \ell_1  T_{w\bar{w}}^g \quad &, \quad
\partial_{\bar w}\alpha =-2\sqrt{2}\ell_1  T_{\bar{w}\bar{w}}^g \nn\\
\partial_w\tilde\alpha =2\sqrt{2}\ell_2  T_{ww}^g \quad &, \quad
\partial_{\bar w}\tilde\alpha =-2\sqrt{2}\ell_2  T_{\bar{w}w}^g\nn
\label{eqn:eom-complex}
\end{align}

These factors of $\sqrt{2}$ can be absorbed into a redefinition of $\ell_1,\ell_2$,
\begin{equation}
    2\sqrt{2} \ell_i \to \ell_i,
    \label{ell-rescale}
\end{equation}
which we will do henceforth.
This redefinition is concomitant with the fact that, in this coordinate system,
\begin{equation}
    \epsilon_{\mu\nu} n^a J^\mu T^\nu_a = - 2 \sqrt{2} (J \bar{T} - \bar{J} T_{w \bar{w}} ), \quad \epsilon_{\mu\nu} \tilde{n}^a \tilde{J}^\mu T^\nu_a = 2 \sqrt{2} (\bar{\tilde{J}} T - \tilde{J} T_{w \bar{w}} ).
    \label{jtbar-flow-reln}
\end{equation}
We now consider two particular theories, where we have concrete expressions for the stress-energy tensor and $U(1)$ currents. 

\subsubsection{Free Scalar}\label{freescalar}

As a first check for the proposal, we apply our formalism to free scalar with undeformed action
\be
{\cal L}_0 = \partial_w\phi \partial_{\bar{w}}\phi  \,.
\ee
By considering a real compact boson, the above theory possesses a $U(1)$ symmetry consisting of constant shifts in field space
\be
U(1): \quad \phi \to \phi + constant \,,
\ee
with associated Noether current
\be
J=-\partial_w\phi \quad , \quad \bar{J}= -\partial_{\bar{w}}\phi \,.
\label{scalarcurrent}
\ee
Note that both components of the current are conserved independently, when imposing the equations of motion of the undeformed theory. They could be taken as independent holomorphic and antiholomorphic currents. However, this cease to be true when the deformation is turned on. We could identify this current with either $J$ or $\tilde{J}$ in our approach.

 For the sake of simplicity, we consider the case of $\ell_2=0$. This identifies \eqref{scalarcurrent} with $J^\nu$. As first step, we promote the global shift symmetry to a local one
 \be
 \phi(\sigma)\to \phi(\sigma) - a(\sigma) \,,
 \label{gaugedscalar}
 \ee
 with $a(\sigma)$ some function of the base space coordinates.
 Then the gauged action
 \be
 S_0[\phi,A_\mu] =\int dw d\bar w (\partial_w\phi+A_w)(\partial_{\bar{w}}\phi+A_{\bar{w}}) \,,
 \ee
 is invariant under a simultaneous shift \eqref{gaugedscalar} and the transformation
 \be
 A_\mu \to A_\mu + \partial_\mu a \,.
 \ee 
We identify the $X^a$ and $\alpha$ fields with the corresponding target space coordinates and gauge transformations respectively. This gives the following  Jacobian and gauge matrices:
 \bea
\left(\begin{array}{cc}\frac{\partial z}{\partial w} & \frac{\partial \bar{z}}{\partial w} \\
\frac{\partial z}{\partial \bar{w}} & \frac{\partial \bar{z}}{\partial \bar{w}}  \end{array}\right)& = &
\left(\begin{array}{cc}1 & \ell_1\left(\frac{\partial z}{\partial w}(\partial\phi-\partial\alpha)+\frac{\partial \bar{z}}{\partial w}(\bar\partial\phi-\bar\partial\alpha)\right) \\
0 & 1 -\ell_1 \left(\frac{\partial z}{\partial \bar{w}}(\partial\phi-\partial\alpha)+\frac{\partial \bar{z}}{\partial \bar w}(\bar\partial\phi-\bar\partial\alpha)\right)\end{array}\right) 
\nn\\
\left(\begin{array}{cc}\frac{\partial \tilde{\alpha}}{\partial w} & \frac{\partial \bar{\alpha}}{\partial w} \\
\frac{\partial \tilde\alpha}{\partial \bar{w}} & \frac{\partial \alpha}{\partial \bar{w}}  \end{array}\right) &=& \left(\begin{array}{cc} 0  & 0 \\
0 & -\ell_1 \left( \frac{\partial z}{\partial \bar{w}}(\partial\phi-\partial\alpha)+\frac{\partial \bar{z}}{\partial \bar{w}}(\bar\partial\phi-\bar\partial\alpha)\right)^2 
\end{array}\right)
\label{scalarJacobian}
\eea

To make the rather abstract ideas of the general procedure as concrete as possible, we will be very explicit in this first example. Since we want to solve for derivatives w.r.t. $z,\bar z$, we need to remember the chain rule in the gauge matrix above.  From the resulting system, we obtain the following solutions
\begin{align}
\left(\begin{array}{cc}\frac{\partial z}{\partial w} & \frac{\partial \bar{z}}{\partial w} \\
\frac{\partial z}{\partial \bar{w}} & \frac{\partial \bar{z}}{\partial \bar{w}}  \end{array}\right) & = \left(\begin{array}{cc} 1 & \ell_1\frac{\partial\phi}{1-\ell_1 \bar\partial\phi} \\
0 & 1-\ell_1 \bar\partial\phi \end{array}\right) \\
\left(\begin{array}{cc}\partial \tilde{\alpha} & \partial \bar{\alpha} \\
\bar\partial \tilde\alpha & \bar\partial \alpha  \end{array}\right) & = \left(\begin{array}{cc} 0 & \ell_1^2\frac{\partial\phi(\bar\partial\phi)^2}{(1-\ell_1 \bar\partial\phi)^2} \\
0 & -\ell_1\frac{(\bar\partial\phi)^2}{1-\ell_1 \bar\partial\phi}  \end{array}\right)
\end{align}
which have to be plugged into \eqref{defaction}. This becomes
\begin{align}
{\cal L} = & \frac{1}{\det(\partial_{w^\mu} z^a)} \Big[(\partial_w z (\partial\phi -\partial\alpha)+\partial_w\bar z (\bar\partial\phi -\bar\partial\alpha))(\partial_{\bar w} z (\partial\phi -\partial\alpha)+\partial_{\bar w}\bar z (\bar\partial\phi -\bar\partial\alpha)) \\
& - \ell_1 (\partial_w z (\partial\phi -\partial\alpha)+\partial_w\bar z (\bar\partial\phi -\bar\partial\alpha)(\partial_{\bar w} z (\partial\phi -\partial\alpha)+\partial_{\bar w}\bar z (\bar\partial\phi -\bar\partial\alpha))^2 \Big]
\end{align}
which nicely simplifies to the following Lagrangian for the deformed theory
\be
{\cal L} = \frac{\partial\phi\bar\partial\phi}{1-\ell_1 \bar\partial\phi} \,,
\ee
which satisfies
\begin{align}
\partial_{\ell_1}{\cal L} &=-J T_{\bar{z}\bar{z}} + \bar{J} T_{z\bar{z}} = \frac{1}{2\sqrt{2}} \epsilon_{\mu\nu}n^a J^\mu T^{\nu}_a \,.
\label{JTflows}
\end{align}
The factor of $2\sqrt{2}$ is correct because of the redefinition of the $\ell$s in \eqref{ell-rescale}.

Alternatively, had we taken $\ell_1=0$ ($J=0$) and deformed with $\ell_2$, the result would read
\be
{\cal L} = \frac{\partial\phi\bar\partial\phi}{1+\ell_2 \partial\phi} \,,
\ee
now satisfying
\begin{align}
\partial_{\ell_2}{\cal L} &=\bar{J} T_{zz} - J T_{\bar{z}z}= \frac{1}{2\sqrt{2}} \epsilon_{\mu\nu}\tilde{n}^a \tilde{J}^\mu T^{\nu}_a \,.
\label{TJflows}
\end{align}  

~ 

To see how the joint flow works, consider the gauged action of two scalar fields
\be
S_0[\phi,\varphi,A_\mu,\tilde{A}_\mu] = \int dw d\bar w (\partial_w\phi+A_w)(\partial_{\bar{w}}\phi+A_{\bar{w}}) +  (\partial_w\varphi+\tilde{A}_w)(\partial_{\bar{w}}\varphi+\tilde{A}_{\bar{w}}) \, ,
\ee
The relevant equations now become
 \bea
\left(\begin{array}{cc}\frac{\partial z}{\partial w} & \frac{\partial \bar{z}}{\partial w} \\
\frac{\partial z}{\partial \bar{w}} & \frac{\partial \bar{z}}{\partial \bar{w}}  \end{array}\right)& =& 
\left(\begin{array}{cc}1+\ell_2 \frac{\partial z^a}{\partial w}(\partial_a\varphi-\partial_a\tilde\alpha) & -\ell_1\frac{\partial z^a}{\partial w}(\partial_a\phi-\partial_a\alpha) \\
-\ell_2 \frac{\partial z^a}{\partial w}(\partial_a\varphi-\partial_a\tilde\alpha) & 1 +\ell_1 \frac{\partial z^a}{\partial \bar{w}}(\partial_a\phi-\partial_a\alpha)\end{array}\right) 
\nn\\
\left(\begin{array}{cc}\frac{\partial \tilde{\alpha}}{\partial w} \\
\frac{\partial \tilde\alpha}{\partial \bar{w}}  \end{array}\right) &=& \left(\begin{array}{cc} \ell_2\left( (\frac{\partial z^a}{\partial w}(\partial_a\phi-\partial_a\alpha))^2+(\frac{\partial z^a}{\partial w}(\partial_a\varphi-\partial_a\tilde\alpha))^2\right)  \\
0 
\end{array}\right)
\\
\left(\begin{array}{cc} \frac{\partial {\alpha}}{\partial w} \\
 \frac{\partial \alpha}{\partial \bar{w}}  \end{array}\right) &=& \left(\begin{array}{cc} 0 \\
 -\ell_1 \left( (\frac{\partial z^a}{\partial \bar{w}}(\partial_a\phi-\partial_a\alpha))^2+(\frac{\partial z^a}{\partial \bar{w}}(\partial_a\varphi-\partial_a\tilde\alpha))^2\right) 
\end{array}\right)\nn
\eea
The same set of steps leads to the jointly deformed action
\be
{\cal L}=\frac{\bar\partial\phi (\partial\phi-\ell_1\partial\varphi\bar\partial\varphi+\ell_2 \partial\phi\partial\varphi)+\bar\partial\varphi(\partial\varphi+\ell_1\partial\phi\bar\partial\varphi-\ell_2(\partial\phi)^2)}{1-\ell_1\bar\partial\phi+\ell_2\partial\varphi+\ell_1\ell_2(\partial\phi\bar\partial\varphi-\bar\partial\phi\partial\varphi)}\,.
\ee
This Lagrangian satisfies both equations (\ref{JTflows}) and (\ref{TJflows}).

\subsubsection{Free Scalar $+$ Free Fermion}

Let us repeat the $J\bar T$ deformation based on the current (\ref{scalarcurrent}) but now in the presence of a Dirac field (decomposed in terms of its left/right moving components). The Fermion contributes to the deformation only through the stress-energy tensor.
The undeformed Lagrangian reads
\be
{\cal L}_0 = \partial_{\bar{w}}\phi\partial_w\phi+ \bar{\psi}\partial_{\bar{w}}\psi + \bar{\chi}\partial_w \chi\,.
\ee
The Jacobian remains that of (\ref{scalarJacobian}). The gauge transformation however receives new contributions from the fermion energy-momentum tensor
\be
\left(\begin{array}{cc} \frac{\partial {\alpha}}{\partial w} \\
 \frac{\partial \alpha}{\partial \bar{w}}  \end{array}\right) = \left(\begin{array}{cc} -\ell_1 \frac{\partial z^a}{\partial {w}}\bar\chi \partial_a \chi  \\
 -\ell_1 \left( (\frac{\partial z^a}{\partial \bar{w}}(\partial_a\phi-\partial_a\alpha))^2+\frac{\partial z^a}{\partial \bar{w}}\bar\chi \partial_a \chi  \right) 
\end{array}\right)\,.
\ee
After solving and inserting into the on-shell action, one finds the deformed Lagrangian
\be
{\cal L}=\frac{\bar\psi \bar\partial \psi+\bar\chi\partial\chi+\ell_1 (\bar\partial\phi)^2\left(\ell_1\bar\psi \bar\partial \psi-\partial\phi\right)+\bar\partial\phi\left(\partial\phi-\ell_1\left(2\bar\psi \bar\partial \psi+\bar\chi\partial\chi\right)\right)}{\left(\ell_1\bar\partial\phi -1\right)^2}\,,
\ee
As a check, this Lagrangian satisfies Equation (\ref{JTflows}) as well.

We conclude by adding that deformations by currents associated to $U(1)$ fermionic phase transformations, as well as combinations of both bosonic and fermionic currents, straightforwardly fit into the presented framework.

\subsection{Adding $T\bar T$ }
We can of course follow the same procedure for the full three-parameter flow including $T \bar{T}$.
The main reason we turned this third term off in the previous section was for simplicity, and a slight conceptual novelty.

 We highlight the limits $\ell_1,\,\ell_2\rightarrow0$ and $\gamma\rightarrow0$ do \textit{not} commute. Non-commutativity of the order in which one deforms the original theory was previously reported in \cite{LeFloch:2019rut}. Considering  $\ell_1,\,\ell_2\rightarrow0$ for finite $\gamma$, the above kernel would give a deformation of the $J\bar J$ type. The $J\bar J$ deformation is marginal and can be dealt with more traditional quantum field theoretical tools. 
In this article, we will also consider the opposite order of limits. That is, to obtain the $T\bar{T}$ deformation alone, it is clear from \eqref{defaction} that we need to take $\ell_{1},\ell_{2} \rightarrow 0$, $\gamma \rightarrow \infty$, keeping $\gamma \ell_{1} \ell_{2} $ fixed. 

We could repeat the steps of previous section to obtain the Lagrangian for the joint $J\bar T$, $T \bar J$ and $T\bar T$ deformed theory. We spare the reader the intermediate steps, and report instead only the starting point and final results.  The computations are straightforward if albeit tedious.

The equations for the Jacobian matrix in complex coordinates are modified to
\bea
\label{diffeo2}
\frac{\partial z}{\partial w}=1-\ell_2 \left. \left(\tilde J_w-\gamma \ell_1 T_{w\bar w}\right)\right|_\alpha\,,\,\,\,\,\,\,\,\,\, \frac{\partial z}{\partial \bar w}=\ell_2 \left. \left(\tilde J_{\bar w}-\gamma \ell_1  T_{\bar w\bar w}\right)\right|_\alpha\,,\nn\\
\frac{\partial \bar{z}}{\partial  w}=-\ell_1 \left. \left( J_w+\gamma \ell_2 T_{w w}\right)\right|_\alpha\,,\,\,\,\,\,\,\,\,\, \frac{\partial \bar z}{\partial \bar w}=1+\ell_1 \left. \left( J_{\bar w}+\gamma \ell_2 T_{\bar w w}\right)\right|_\alpha\,,
\eea
while the last two lines of \eqref{complexdiff} remain the same. We turn to the Dirac fermion for concreteness.
Here, we have rescaled $\ell_1,\ell_2$ to capture rogue factors of $2\sqrt{2}$.

\subsubsection{Free Fermion}

The undeformed Lagrangian for a free Dirac fermion reads
\be
{\cal L}_0 =  \bar{\psi}\partial_{\bar{w}}\psi + \bar{\chi}\partial_w \chi,
\label{freeferm}
\ee
This theory contains two $U(1)$ currents which act independently on the $\psi$ and $\chi$ fields as a phase transformations: $\psi\rightarrow e^{-i\alpha} \psi$ and $\chi\rightarrow e^{-i\tilde{\alpha}} \chi$. The complex conjugate spinors transform with opposite phases. The undeformed action coupled to the gauge fields is given by,
\bea
S_0[\psi,\bar{\psi},\chi,\bar{\chi},A_\mu, \tilde{A}_\mu]=\int dw d \bar{w}\big(\bar{\psi}(\pa_{\bar{w}}+A_{\bar{w}})\psi+\bar{\chi}(\pa_w+\tilde{A}_{w})\chi\big),
\eea
with associated currents
\begin{align}
J_L = -\bar{\psi}\psi \quad &, \quad \bar{J}_L=0 \nn\\
J_R=0 \quad &, \quad \bar{J}_R = -\bar{\chi}\chi
\end{align}
Note that ($J_R$)$J_L$ is naturally a (anti-)holomorphic current. In contrast to the example of the compact boson, the currents associated to these symmetries transform as scalars under the coordinate transformation and are left invariant by the $U(1)$ transformations. This in fact simplifies the computations.  
This property holds along the flow.\\
We consider the deformation given by turning on both parameters $\ell_1$ and $\ell_2$ to arbitrary values. This identifies $J_L$ with $J$ and $J_R$ with $\tilde{J}$.

Equations \eqref{diffeo2} and \eqref{defaction} jointly give
\be
{\cal L} =  \bar{\psi}\bar\partial\psi + \bar{\chi}\partial \chi + \ell_1 \bar\psi \psi (\bar\chi \bar\partial \chi) - \ell_2 \bar\chi \chi (\bar\psi \partial\psi)-\gamma \ell_1\ell_2 \left( \bar\psi \partial\psi\bar\chi \bar\partial \chi-\bar\psi \bar\partial\psi\bar\chi \partial \chi\right)
\ee
By identifying $\lambda=\gamma\ell_1\ell_2$, it is not difficult to check the above Lagrangian satisfies the following joint flow equations
\begin{align}
\partial_{\ell_1}{\cal L} &=-J T_{\bar{z}\bar{z}} + \bar{J} T_{z\bar{z}}=
\frac{1}{2\sqrt{2}}\epsilon_{\mu\nu}n^a J^\nu T^{\mu}_a\\
\partial_{\ell_2}{\cal L} &=\bar{\tilde J} T_{zz} - \tilde J T_{\bar{z}z}=
\frac{1}{2\sqrt{2}}\epsilon_{\mu\nu}\tilde{n}^a \tilde{J}^\nu T^{\mu}_a\\
\partial_{\lambda}{\cal L}&= -T_{zz}T_{\bar z\bar z}+T_{z\bar z}T_{\bar z  z} =\frac{1}{8} \epsilon_{\mu\nu}\epsilon^{ab}T^\mu_a T^\nu_b
\end{align}
for any value of $\ell_1$, $\ell_2$. It is worth mentioning that, for the deformed theory, both the left and right currents acquire a non-chiral and chiral components respectively, $\bar{J}_L= \ell_2 \bar\psi \psi \bar \chi \chi$ and $J_R= -\ell_1 \bar\psi \psi \bar \chi \chi$. However, by the Grassmanian nature of the fermionic fields, both vanish when multiplied by any component of the stress-energy tensor. Thus, they  do not appear in the flow equation. By setting different parameters to zero, we recover multiple cases of interest. For example, setting $\ell_2=0$ we recover the example of the Dirac fermion under the $J\bar{T}$ deformation, considered in \cite{Guica:2017lia}. It was also noted there the $J$ current remained chiral all along the flow. Taking instead $\gamma\rightarrow 0$, one recovers the $J\bar{T}+T\bar{J}$ deformation discussed in section \ref{JTbar+TJbar}.

\subsubsection{Free Boson}

We conclude our classical check of the kernel with one final example.  We return to the free massless boson and consider its $T\bar T+J \bar T$ deformation. By choosing $J_\mu$ to be the current associated to constant shifts of $\phi$, as in \eqref{scalarcurrent}, and setting $\tilde J_\mu=0$ identically, we end up with the following deformed Lagrangian 
\be
{\cal L}=\frac{1-\ell_1\bar\partial\phi -\sqrt{\left(1-\ell_1 \bar\partial\phi\right)^2-4 \lambda \bar\partial\phi \partial\phi  }}{2\lambda}\,,
\ee
where, again, we introduced $\lambda=\gamma\ell_1\ell_2$. The above Lagrangian nicely satisfies the flow equations
\bea
\partial_{\ell_1}{\cal L}&=&  \frac{1}{2\sqrt{2}}\epsilon_{\mu\nu}n^a J^\mu T^\nu_a\\
\partial_{\lambda}{\cal L}&=& \frac18   \epsilon_{\mu\nu}\epsilon^{ab}T^{\mu}_aT^\nu_b\,.
\eea

\section{The Quantum Partition Function} \label{quantum}

We now probe our proposed kernel's validity at the fully quantum level. To do so, we wish to explicitly compute the path integral over base space torus geometries and gauge connections: 
\begin{eqnarray}
Z_{\ell_{1},\ell_{2}}[f,B,\tilde{B}] & = & \int\frac{DeDY}{\text{vol(diff)}}\frac{DA D\alpha}{\text{vol}(G)}\frac{D\tilde{A} D\tilde{\alpha}}{\text{vol}(\tilde{G})}e^{-\frac{1}{\ell_{1}}\int_{T_{BS}^2} \tilde{n}_{a}(f-e)^{a}\wedge(B-A)-\alpha \tilde{n}_{a}de^{a}+\tilde{n}_{a}Y^{a}dA} \nn \\
 &  & \qquad \qquad \times e^{-\frac{1}{\ell_{2}}\int_{T_{BS}^2} n_{a}(f-e)^{a}\wedge(\tilde{B}-\tilde{A})-\tilde{\alpha} n_{a}de^{a}+n_{a}Y^{a}d\tilde{A}}\nn\\
 &  & \qquad \qquad \times e^{\gamma \int_{T_{BS}^2} (B-A) \wedge (\tilde{B}-\tilde{A}) + \alpha d\tilde{A} - \tilde{\alpha} dA }\ Z_{0}[e,A,\tilde{A}],
 \label{Zbyparts}
\end{eqnarray}
where in the above expression we have imposed \eqref{eqn:kernel-reqts} and  rewritten the action after having performed an integration by parts to make manifest the constraints imposed by the $Y^a$ and $\alpha, \tilde{\alpha}$ integrals. 

 First of all, we introduce the arbitrary length scale $s$ which determines the integration domain of base space coordinates $\sigma^\mu$, {\it i.e.} $s=\int d\sigma^1=\int d\sigma^2$. Of course, at the end of the computation, nothing will depend on the particular choice of $s$. Nonetheless, it turns out to be useful to keep track of dimensions.

We take the target space fields to be
\begin{align}
    f_\mu^a = \frac{1}{s} L_\mu^a \quad , \quad 
    B_\mu =  \frac{1}{s} b_\mu \quad , \quad 
    \tilde{B}_\mu = \frac{1}{s} \tilde{b}_\mu.
    \label{eqn:ts-fields-t2}
\end{align}
These are the natural choices given that they need to be flat.
The $L^a$s naturally parametrize the size and orientation of the two cycles of the torus, and the $b,\tilde{b}$ are the holonomies.

In the pure $T\bar{T}$ case, \cite{Dubovsky:2018bmo} found their path integral over gravitational degrees of freedom localized to an integral solely over global modes. We find that both the gravitational and the gauge degrees of freedom localize similarly in our case. We follow \cite{Dubovsky:2018bmo} closely. 

Before delving into technical details, let us outline the three main
steps in our computation. 
\begin{itemize}
\item The first is standard: we need to avoid overcounting diffeomorphism and $U(1)$ gauge equivalent configurations. For diffeomorphisms, we accomplish this by writing a general vielbein in terms of a Weyl rescaling, a local Lorentz transformation (an $SO(2)$ rotation in our Euclidean setup) and a diffeomorphism of some fixed reference vielbein: $e^{a}(\sigma)=\left(e^{\Omega(\sigma)}\left(e^{\epsilon\phi(\sigma)}\right)_{b}^{a}\hat{e}^{b}\right)^{\xi}$
. The change of variables from $e$ to $\Omega,\phi$ and $\xi$ is
accompanied by an important Jacobian, the (diffeomorphism) Fadeev-Popov determinant. We can import the result of \cite{Dubovsky:2018bmo} here.
Similarly, we can decompose the gauge field as 
\begin{equation}
  A = A_{H}+ dg(\sigma) + \star d\chi(\sigma)  = (A_{H,\mu} + \partial_{\mu} g + g_{\mu\nu} \epsilon^{\nu\rho} \partial_{\rho} \chi) d\sigma^{\mu}, \quad dA_{H} = \star  dA_{H} = 0.
\end{equation}
where the gauge-invariant $A_{H}$ encode the holonomies as
\begin{equation}
  A_{H} = \frac{1}{s} h_{\mu} d\sigma^{\mu}, \quad \oint A_{H,1} d\sigma^{1} = h_{1},\quad \oint A_{H,2} d\sigma^{2} = h_{2}.
  \label{eqn:A-holonomy-piece}
\end{equation}
The Jacobian for change of variables from $A$ to $A_{H},g$ and  $\chi$ gives the other $U(1)$ Fadeev-Popov determinant. The same obviously holds for $\tilde{A}$.

\item The second step will be to understand precisely the constraints arising from the integrals over the $Y$'s and $\alpha ,\tilde{\alpha}$. Beyond their important role in localizing the path integral to minisuperspace, we determine the additional functional determinants they contribute to the path integral.

\item Finally, we will need to treat field zero modes with great care. \cite{Dubovsky:2018bmo} already showed how the zero-mode integral for the $Y^a$ gave an important factor of the area of the \textit{target} space torus. We will find that the range of integration for the analogous holonomies of the gauge fields $A, \tilde{A}$ has equally important consequences.   

\end{itemize}

\subsection{A Note on the Gauge Symmetries} \label{gauge-unwrap}
Before we begin the main computation, we need to elucidate a slightly subtle point about the diffeomorphism and $U(1)$ gauge symmetries.
We will find that these two gauge symmetries consist of only those transformations connected to the identity. This has an important impact on the range of integration for the moduli and holonomies.

First, consider the $U(1)$ gauge symmetry of $A$, whose transformations are given by \eqref{eqn:u1-g-tr-int}, reproduced here for readability,
\begin{equation}
    \delta A_\mu = \partial_\mu g, \quad \delta \alpha = g.
    \label{eqn:u1-g-tr}
\end{equation}
The important thing is that the symmetry depends on the ability to absorb the gauge transformation into $\alpha$.
We stress this point because $\alpha$ is single-valued on the torus. Therefore, this is a symmetry iff $g$ is also single-valued.

Now, consider the gauge field configuration
\begin{equation}
    A_\mu d\sigma^\mu = \frac{2\pi}{q} \frac{d \sigma^1}{s} = d \left( \frac{2\pi}{q} \frac{\sigma^1}{s} \right),
    \label{eqn:u1-triv-hol-eg}
\end{equation}
where $q$ is the unit of fundamental $U(1)$ charge and serves as the inverse radius of the $U(1) \cong S^1$. In usual $U(1)$ gauge theories, since the gauge group is compact, the condition on the gauge transformations is
\begin{equation}
    e^{i q \, g(s)} = e^{i q \, g(0)}.
\end{equation}
Because of this, $g$ itself need not be single-valued,
\begin{equation}
    e^{i q \, g(s)} = e^{i q \, g(0)} \quad \Rightarrow \quad g(s) - g(0) \in \frac{2\pi}{q} \mathbb{Z}.
    \label{eqn:winding-g-tr}
\end{equation}

With this condition, the gauge field configuration in \eqref{eqn:u1-triv-hol-eg} is gauge-equivalent to $0$, and therefore the holonomies are compact --- as they should be, given the compactness of the gauge group.
However, in our case, we have 
\begin{equation}
    g(s) = g(0) \quad \Rightarrow \quad A= \frac{2\pi}{q} \frac{d \sigma^1}{s} \not\sim  A = 0 \quad \Rightarrow \oint A \in \mathbb{R}.
    \label{eqn:hol-non-cpct-cond}
\end{equation}
In other words, the restriction to gauge transformations connected to the identity cause the holonomies to be valued not in the group itself but in its universal cover --- which is $\mathbb{R}$ for $U(1)$. The story for the other $U(1)$ is of course identical.

The analogous restriction on the diffeomorphisms imply the moduli $\tau$ are valued not in the fundamental domain but in the entire upper half plane; see \cite{Dubovsky:2018bmo} for a more detailed discussion.

\subsection{Path Integral Measures and Fadeev-Popov Determinants} \label{localisation}

We finally begin our computation, focusing on relevant path integral ``measures''.\footnote{Obviously, there is no sense in which we may hope to truly \textit{define} the path integral measures. Instead, we are simply making explicit our bookkeeping.} Our discussion parallels Polchinski's original computation of the Polyakov string torus path integral \cite{PolchinskiTorus}.

For an $n$-dimensional manifold with coordinates $x^{i}$, the invariant measure is
\begin{equation}
  d^{n} x \sqrt{g(x)}, \quad \text{where} \quad ds^{2} = g_{ij}(x) dx^{i}  dx^{j};
  \label{eqn:measure-n-dim}
\end{equation}
in other words, the measure depends on $g_{ij}$, which in turn can be defined through the inner product on small variations of the coordinates. In infinite dimensions, we cannot be so explicit. Rather, we define it implicitly via fixing the value of a Gaussian path integral. 
The first step therefore requires defining an inner product on infinitesimal variations of the fields living in the tangent space at a given point on the field space manifold.
We choose
\begin{align}
  (\delta e, \delta e)_{e} &= s^{-2} \int \delta e^{a} \wedge \star  \delta e_{a} = s^{-2} \int (\det e) \delta e_{\mu}^{a} \delta e^{\mu}_{a} d^2 \sigma \nonumber\\
  (\delta Y, \delta Y)_{e} &= s^{-2} \int \delta Y^{a} \wedge \star  \delta Y^{a} = s^{-2}\int (\det e) \delta Y^a \delta Y_a d^2 \sigma \nonumber\\
  (\delta A, \delta A)_{e} &= s^{-2} \int \delta A \wedge \star  \delta A = s^{-2}\int (\det e) \delta A_{\mu} \delta A^{\mu} d^2 \sigma \nonumber\\
  (\delta \alpha, \delta \alpha)_{e} &= s^{-2} \int \delta \alpha \wedge \star  \delta \alpha= s^{-2}\int (\det e) \delta \alpha \delta \alpha d^2 \sigma \,.
  \label{eqn:field-ips}
\end{align}
The factors of $s$ have been arbitrarily inserted so that the `field-space metric' is dimensionless. In other words, it is an arbitrary length scale chosen to cancel the factors of length arising from the integration measure $d^2 \sigma$. These inner-products are diffeomorphism invariant and depend on the base space metric variables in the form of the vielbeins $e^{a}_{\mu}$ used to raise and lower indices. In particular, this implies that the path integral measure for the vielbeins will be non-linear. 
We have not explicitly written the inner products for $\tilde{A}$ and $\tilde{\alpha}$, since they are the same as the last two equations in \eqref{eqn:field-ips}.

Denoting by $\Psi $ whatever field we are interested in, we then implicitly define the measure at a given point in field space by requiring

\begin{equation}
    \int D \delta\Psi e^{-\frac{1}{2}(\delta \Psi, \delta \Psi)} = 1
\end{equation}

To compare with the familiar finite dimensional case, this would give:
\begin{equation}
    \int D \delta x e^{-\frac{1}{2} g(x)_{i j}\delta x^{i}\delta x^{i}} = \sqrt{\frac{\pi^n}{det(g(x))}} =1 
\end{equation}
which would define the measure $Dx= \sqrt{\frac{det(g(x))}{\pi^n}}d^n x$

In the infinite-dimensional case, we can absorb factors such as $\pi$ via local counterterms, so those will not be of much importance \cite{PolchinskiTorus}.\footnote{We will need to be more careful when restricting an integral to non-zero modes, as the above statement about local counter-terms no longer holds. We will have to keep track of those (as we will see in expressions like $\det'(2\pi)$, where $'$ denotes exclusion of zero modes).} This might all seem overkill at first sight. In fact, it greatly simplifies the the computation of the Fadeev-Popov determinants and informs our treatment of the zero-modes.

\subsubsection{Measures and Fadeev-Popov Determinant for the $U(1)$ gauge fields}
In this subsection, we will calculate the Faddeev-Popov determinants needed to address the gauge-invariance of the path integral.

The Hodge decomposition theorem guarantees we can decompose any one-form $A$ as
\begin{equation}
  A =  A_{H} + dg(\sigma) + \star d\chi(\sigma) = ( A_{H,\mu}+ \partial_{\mu} g + g_{\mu\nu} \epsilon^{\nu\rho} \partial_{\rho} \chi  ) d\sigma^{\mu}, \quad dA_{H} = \star dA_{H} = 0.
  \label{eqn:A-decomp}
\end{equation}
Here, we are taking $g,\chi$  to contain no constant pieces (zero-modes), since those would not contribute to $A$.
All zero-modes of $A$ are contained in $A_{H}$. These encode the holonomies and can be written in terms of them as
\begin{equation}
  A_{H} = \frac{1}{s} h_{\mu} d\sigma^{\mu}, \quad \oint A_{H,1} d\sigma^{1} = h_{1},\quad \oint A_{H,2} d\sigma^{2} = h_{2}.
  \label{eqn:A-holonomy-piece}
\end{equation}
For reasons discussed in section \ref{gauge-unwrap}, the holonomies take values in the universal cover of $U(1)$, i.e. the full real line $\mathbb{R}$.

We find the Jacobian for this change of variables via 

\begin{equation}
1 = \int D\delta A e^{-\frac{1}{2}(\delta A,\delta A)} = J_{U(1)} \int D\delta A_{H} D\delta g' D\delta{\chi}'  e^{- \frac{1}{2} (\delta A,\delta A)}
\end{equation}

where we need to express $(\delta A,\delta A)$ in terms of $\delta A_{H}, \delta g'$ and $\delta{\chi}'$ in the last equality. 
Straightforward algebra gives
\begin{equation}
  (\delta A, \delta A) =  (\delta g, -\Box \delta g) + (\delta \chi, -\Box \delta \chi) + \frac{1}{s^2}\int (\det e) \delta A_{H,\mu} \delta A_{H}^{\mu}.
  \label{eqn:A-ip-decomp}
\end{equation}
from which we conclude
\begin{equation}
  DA = \left( D'g D'\chi \det'( -\Box) \right) DA_{H},
  \label{eqn:A-measure-decomp}
\end{equation}
with the various measures defined using the inner products appearing in \eqref{eqn:A-ip-decomp}. The primes denote exclusion of zero-modes. Note the dependence on the base space vielbeins via the functional determinant of the Laplacian $\delta^{ab} e^{\mu}_a e^{\nu}_b \nabla_{\mu} \nabla_{\nu }= \Box $.

$g$ in \eqref{eqn:A-decomp} really paramterizes the pure-gauge direction in $A$. We have already shown the kernel is gauge-invariant. Hence, it does not depend on $g$. We therefore can pull it out of the rest of the integral and need simply evaluate the ratio 
\begin{equation}
 \frac{\int Dg'}{\text{vol}(G)}\,.
\end{equation}
 
 This ratio is not quite unity, because of the exclusion of zero modes in the numerator. Indeed, $\text{vol}(G)=\int Dg = \int D\bar{g} \int Dg'$
 where we split up a general group element $g$ into a sum of zero- and non-zero-mode pieces
\begin{equation}
  g = \bar{g} + g'\,, \quad d \bar{g} = 0\,,\quad ( \delta g',\delta \bar{g}) = 0\,, 
  \label{eqn:g-decomp}
\end{equation}
the zero-mode piece $\bar{g}$ is compact because of the compactness of the group $U(1)$
\begin{equation}
  \bar{g} \in \left[ 0, \frac{2\pi}{q} \right),
  \label{eqn:g-bar-restriction}
\end{equation}
and the inner product decomposes nicely as
\begin{equation}
  (\delta g, \delta g) = (\delta \bar{g}, \delta \bar{g}) + (\delta g', \delta g').
  \label{eqn:g-ip-decomp}
\end{equation}
which shows the zero mode and non-zero mode pieces are orthogonal realtive to the inner product. 
Using these two facts we have that
\begin{align}
  \int Dg &= \int D \bar{g} \int Dg' \nonumber\\
  &= \frac{\sqrt{\bar{\mathcal{A}}}}{s} \int d \bar{g} \int Dg' \nonumber\\
  &=  \frac{\sqrt{\bar{\mathcal{A}}}}{s} \frac{2\pi}{q} \int Dg',
  \label{eqn:Dg-measure-ratio}
\end{align}
$\bar{\mathcal{A}}$ denotes, as in \cite{Dubovsky:2018bmo} the proper area of the base space torus. 
This gives us the explicit ratio:
 \begin{equation}
     \frac{\int Dg'}{\text{vol}(G)} = \frac{s}{\sqrt{\bar{\mathcal{A}}}} \frac{q}{2\pi}\,.
     \label{eqn:gauge-ratio}
 \end{equation}
 Of course, all the above steps are identical for the second gauge field $\tilde{A}$ and the quotient by $\text{vol}(\tilde{G})$. 
 
 \subsubsection{Measures and Fadeev-Popov Determinant for the Vielbeins}
 This section is short. All the hard work has already been done in \cite{Dubovsky:2018bmo} and we can straightforwardly import their results. 

To be precise, recall that any vielbein $e^a$ on the torus may be written as 
\begin{equation}
  e^{a} = \left( e^{\Omega} e^{\phi \varepsilon^{a}_{\ b}} \hat{e}^{b} (\tau) \right)^{\xi},
  \label{eqn:e-decomp}
\end{equation}
where $(\dots)^{\xi}$ means a finite diffeomorphism generated by the vector field $\xi$ and the canonical unit-torus veilbeins $\hat{e}^a$ are given by
\begin{equation}
  \hat{e}^{1} (\tau') = d\sigma^{1} + \tau'_{1} d\sigma^{2}, \quad \hat{e}^{2} (\tau') = \tau'_{2} d\sigma^{2}.
  \label{eqn:unit-t2-e}
\end{equation}
The decomposition of the measure in this case is more involved. In fact, we will only need the Jacobian satisfying the constraints imposed by the Lagrange multipliers.  On that constraint surface, we will see all non-zero modes of the vielbeins vanish. We quote the answer \cite{Dubovsky:2018bmo} found for later convenience:
\begin{equation}
  De = D\xi' D\Omega D\phi d^{2} \tau' J_{\text{diffs}} \overset{\text{constraints}}{=} D\xi' D\Omega D\phi d^{2} \tau' \frac{\bar{\mathcal{A}}}{s^2 (\tau')_{2}^{2}} \det' (- \Box).
  \label{eqn:e-measure-decomp}
\end{equation}

~

Note that in this decomposition we have also excluded $\xi$ zero-modes, as these do not change the vielbein and would render our paramterization redundant (see \cite{PolchinskiTorus} for more details). Since our kernel respects base space diffeomorphism invariance, the $D\xi'$ similarly decouples. As for the $U(1)$'s , we again need to be careful about the ratio $\int D\xi' / \text{vol(diff)} $, which was also found in  \cite{Dubovsky:2018bmo}
\begin{equation}
  \frac{\int D'\xi}{\text{vol(diff)}} = \frac{ s^2 }{\bar{\mathcal{A}}^2}\,.
  \label{eqn:xi-ints-ratio}
\end{equation}
This completes our necessary list of ingredients to proceed to the constraint integrals.

\subsubsection{The Constraint Integrals} \label{ssec:constraints} 
The next step is to perform the integrals over the compensator fields $\alpha,\tilde{\alpha},Y^{a}$.
From \eqref{Zbyparts}, all three of these clearly impose $\delta$ functions. We need to evaluate the additional functional determinant prefactors they contribute to the path integral.
We begin with the $\alpha$ and $\tilde{\alpha}$ integrals, rewriting the relevant part of their action in terms of their associated inner product: 
\begin{equation}
\int D\alpha D\tilde{\alpha} e^{-(S_{\alpha}+S_{\tilde{\alpha}})}=\int D\alpha D\tilde{\alpha} e^{-(\alpha',-\frac{s^2}{\ell_{1}}\star \tilde{n}_a de^a + s^2 \gamma \star d\tilde{A})-(\tilde{\alpha}',-\frac{s^2}{\ell_{2}}\star n_a de^a - s^2 \gamma \star dA)}
  \label{eqn:s-alpha}
\end{equation}
 Integration by parts shows only the non-zero modes of $\alpha$ and $\tilde{\alpha}$ contribute to exponent. Indeed, we can mimick our treatment of $\text{vol}(G)= \int Dg $ and split $\alpha$ into its zero mode and orthogonal non-zero mode piece $ \alpha = \bar{ \alpha } + \alpha'$ (and similarly for $\tilde{\alpha}$). Further remember that the $\bar{\alpha},\bar{\tilde{\alpha}} \in U(1)$ are compact. The integrals in \eqref{eqn:s-alpha} thus become

\begin{align}
  \int_{0}^{2\pi / q}d \bar{\alpha} \int_{0}^{2\pi / \tilde{q}} d \bar{\tilde{\alpha}} \int D\alpha' D \tilde{\alpha}' e^{-(S_{\alpha}+S_{\tilde{\alpha}})} = \frac{\bar{\mathcal{A}}}{s^{2}} \frac{(2\pi)^{2}}{q \tilde{q}} &\delta' \left( \frac{-s^{2}}{2\pi\ell_{1}} \tilde{n}_{a} \star de^{a} + \frac{s^2}{2\pi} \gamma \star d \tilde{A} \right)\nn\\ 
  &\delta' \left( \frac{-s^{2}}{2\pi\ell_{2}} n_{a} \star  de^{a} - \frac{s^2}{2\pi} \gamma \star dA \right).
  \label{eqn:alpha-int}
\end{align}
where the $'$ reminds us of the exclusion of zero-modes. In all these expressions, we should really be writing the vielbeins in terms of $\Omega, \phi$ and $\hat{e}^a$ (diff. invariance tells us we can ignore $\xi$), but have avoided doing so to avoid cluttering the notation even further.

We now turn to the $Y^a$ integrals. The action for $Y$ reads
\begin{equation}
  S_{Y} = \int Y^{'a} \left( \frac{1}{\ell_{1}} \tilde{n}_{a} dA + \frac{1}{\ell_{2}} n_{a} d\tilde{A} \right) d^2 \sigma = \left( Y^{'a}, \frac{s^{2}}{\ell_{1}} \tilde{n}_{a} \star dA + \frac{s^{2}}{\ell_{2}} n_{a} \star  d \tilde{A} \right).
  \label{eqn:s-y}
\end{equation}
where we have again decomposed $Y^a=\bar{Y}^a+Y'^a$ into its zero- and non-zero mode contributions. 
Using the approriate measures for the $Y$, the integral becomes
\begin{equation}
  \int DY e^{-S_{Y}} = \left( \int d^2 \bar{Y} \frac{\bar{\mathcal{A}}}{s^2} \right) \delta' \left( \frac{s^{2}}{\ell_{1}} \tilde{n}_{a} \star dA + \frac{s^{2}}{\ell_{2}} n_{a} \star  d \tilde{A} \right) = \frac{\mathcal{A} \bar{\mathcal{A}}}{s^{2}} \delta' \left( \frac{s^{2}}{2\pi\ell_{1}} \tilde{n}_{a} \star dA + \frac{s^{2}}{2\pi\ell_{2}} n_{a} \star  d \tilde{A} \right).
  \label{eqn:Y-int}
\end{equation}
Similarly, these equations should be understood for $dA = d\star d\chi $.

These three $\delta$ functions are somewhat cumbersome to work with. They involve different components and combinations of the gauge fields and vielbeins. We can tease these apart using
\begin{equation}
  \left(
  \begin{matrix}
    \frac{-s^{2}}{2\pi\ell_{1}} \tilde{n}_{a} \star de^{a} + \frac{s^2}{2\pi} \gamma \star d\tilde{A} \\
    \frac{-s^{2}}{2\pi\ell_{2}} n_{a} \star de^{a} - \frac{s^2}{2\pi} \gamma \star dA\\
    \frac{s^{2}}{2\pi\ell_{1}} \tilde{n}_{a} \star dA + \frac{s^{2}}{\ell_{2}} n_{a} \star d \tilde{A}\\
  \end{matrix} \right) = \frac{s^{2}}{2\pi} \left(
  \begin{matrix}
    \frac{-1}{\ell_{1}} \tilde{n}_{a} & 0  &\gamma \\
    \frac{-1}{\ell_{2}} n_{a} & -\gamma  &0\\
    0  & \frac{1}{\ell_{1}} \tilde{n}_{a}  & \frac{1}{\ell_{2}} n_{a}\\
  \end{matrix}
  \right)  \left(
  \begin{matrix}
    \star de^{a} \\
    \star dA\\
    \star d \tilde{A}\\
  \end{matrix}
  \right) 
  \label{eqn:delta-fn-rot-mat}
\end{equation}
Note that this is a $4 \times 4$ matrix. This means that the delta functions can be rewritten, using the defining properties of the $n$'s, as 
\begin{align}
  \delta' \left( \frac{-s^{2}}{2\pi\ell_{1}} \tilde{n}_{a} \star de^{a} + \frac{s^2}{2\pi} \gamma \star d \tilde{A} \right) &\delta' \left( \frac{-s^{2}}{2\pi\ell_{2}} n_{a} \star  de^{a} - \frac{s^2}{2\pi} \star dA \right)  \delta' \left( \frac{s^{2}}{2\pi\ell_{1}} \tilde{n}_{a} \star dA + \frac{s^{2}}{2\pi\ell_{2}} n_{a} \star  d \tilde{A} \right) \nonumber\\
  &= \left(\det' \left( \frac{s^{8}}{4(2\pi\ell_{1})^{2} (2\pi\ell_{2})^{2}} \right) \right)^{-1}  \delta(\star de^{a}) \delta(\star dA) \delta(\star d \tilde{A}),
  \label{eqn:delta-fn-trans-det}
\end{align}
where the factor of $4$ is the determinant of the $2 \times 2$ matrix $(\tilde{n}_a,\ n_a)$.

The determinant excludes zero-modes. Finally, write the delta-function constraints explicitly in terms of our new integration variables:
\begin{align}
  \star dA &= \Box \chi \nonumber\\
  \star de^{a} &= \star d \Omega \hat{e}^{a} + \star d\phi \varepsilon^{a}_{\ b} \hat{e}^{a}.
  \label{eqn:delta-fn-arg-dec}
\end{align}
Again using results from \cite{Dubovsky:2018bmo} for the vielbein sector, this gives us
\begin{equation}
  \delta(\star de^{a}) \delta(\star dA) \delta(\star d \tilde{A}) = \frac{1}{(\det' -\Box)^{3}} \delta' (\chi) \delta' (\tilde{\chi}) \delta' (\Omega) \delta' (\phi).
  \label{eqn:delta-fn-final}
\end{equation}
Notice that the $\det'(-\Box)$ factors exactly cancel the ones coming from the Fadeev-Popov determinants, much as in \cite{Dubovsky:2018bmo}.

\subsubsection{Final Answer}
Doing the constraint integrals exposed the inner workings of the path integral localization to zero-modes. We need only two more ingredients before we can put it all together. First, since the global scale and rotation are not fixed, we need the measure for them.
Denoting the zero-modes of $\Omega$, $\phi$ as $\bar{\Omega}$,$\bar{\phi}$, the relevant part of the measure is
\begin{equation}
  D \bar{\Omega} D \bar{\phi} = d \bar{\Omega} d \bar{\phi} \frac{\bar{{\cal A}}}{s^{2}}.
  \label{eqn:omega-phi-0-mode-measure}
\end{equation}
Secondly, we need to deal with the $\det'$ of a constant in \eqref{eqn:delta-fn-trans-det}.
For this, we use the fact that the $\det$ of a constant is merely an addition to the cosmological constant and can therefore be absorbed into a choice of counterterm in $Z_{0}$, so that we can write
\begin{equation}
  \det' c = \frac{1}{c}.
  \label{eqn:det-p-const}
\end{equation}

The measure for the holonomies, using \eqref{eqn:A-ip-decomp} and the localisation to constant vielbeins, reads
\begin{equation}
  DA_{H} = \frac{dh_1 dh_2}{s^2}\frac{\bar{\mathcal{A}}}{s^2}\frac{1}{det[\bar{e}]}=\frac{dh_1 dh_2}{s^2}\,.
  \label{eqn:hol-measure}
\end{equation}

Finally, we parametrize the vielbeins in terms of the length vectors of the two cycles as
\begin{equation}
    f_\mu^a = \frac{1}{s} L_\mu^a, \quad e_\mu^a = \frac{1}{s} \bar{L}_\mu^a.
    \label{eqn:f-e-param}
\end{equation}
which in turn can be parametrized in the following way
\begin{align}
\bar{L}^1_1 &= s e^{\Omega} \cos\phi\\
\bar{L}^1_2 &= s e^{\Omega} \sin\phi\\
\bar{L}^2_1 &= s e^{\Omega} (\tau'_1 \cos\phi- \tau'_2\sin\phi)\\
\bar{L}^2_2 &= s e^{\Omega} (\tau'_1\sin\phi+\tau'_2 \cos\phi)
\end{align}
and analogously for the target variables. It is easy to check that 
\be
\bar{\cal A}\frac{d\Omega d\phi d^2\tau'}{(\tau')_{2}^{2}}=\frac{d^2 \bar{L}^1d^2\bar{L}^2}{\bar{{\cal A}}} = \frac{d^2\bar{L}^1d^2\tau'}{\tau'_2}
\ee

We may now plug these things into the full path integral. We spare the reader the details. Instead, we simply note a few important cancellations:
\begin{enumerate}
  \item The integral over the zero-modes of $\alpha,\tilde{\alpha},Y$ exactly cancel the parts the original volumes of gauge groups that failed to cancel in \eqref{eqn:gauge-ratio}.
  \item The scalar Laplacian determinants that arise from the gauge-fixing cancel those from the delta-functions.
\end{enumerate}
After what have admittedly been many steps, the full partition function simplifies to
\begin{align} \label{torusfinal}
Z_{\ell_1 , \ell_{2}} [f, b, \tilde{b}] = \frac{{\cal A}}{4(2\pi)^{4} \ell_{1}^{2} \ell_{2}^{2}} \int  &d^{2} h d^{2} \tilde{h} \frac{d^2 \bar{L}^1d^2\tau'}{\tau'_2} \nn\\ &e^{-\left( \frac{1}{\ell_1}  \tilde{n}_a (L-\bar{L})_\mu^a (b-h)_\nu + \frac{1}{\ell_2} n_a (L-\bar{L})_\mu^a (\tilde{b} - \tilde{h})_\nu - \gamma (b-h)_\mu (\tilde{b} - \tilde{h})_\nu \right)\epsilon^{\mu \nu } } Z_0[e,h,\tilde{h}]
\end{align}
which is $s$-independent and dimensionless, as it should be. This is one of the cornerstone results of this paper. 
In writing this expression, we have, in the interest of aesthetics, shifted $h, \tilde{h}$ by $b,\tilde{b}$.

It is trivial to check that this final simplified form of the partition function satisfies the diffusion-type equations \eqref{diffusion},
\begin{equation}
    \partial_{\ell_1} \mathcal{Z}_{\ell_1,\ell_2} = - \epsilon_{\mu\nu} \partial_{b_\mu} n^a \partial_{L_\nu^a}  \mathcal{Z}_{\ell_1,\ell_2} \equiv - n^a \partial_b \wedge \partial_{L^a} \mathcal{Z}_{\ell_1,\ell_2}, \quad \mathcal{Z}_{\ell_1,\ell_2} \equiv \frac{Z_{\ell_1,\ell_2}}{\mathcal{A}},
    \label{eqn:global-flow-text}
\end{equation}
and a similar equation for $\ell_2$, where we have to write $\gamma$ in terms of $\lambda$ using \eqref{eqn:lambda} to show this.
While the manipulations of the following sections can only be done for very special currents, the results obtained so far hold for any pair of $U(1)$ symmetries. 

\subsection{The Anomalous Case} \label{ssec:anom}
In this section, we point out the somewhat surprising fact that the path integral \eqref{defpartition} reduces to the finite-dimensional integral \eqref{torusfinal} even in the case where the seed theory has a `t Hooft anomaly.
The reason this is surprising is that the existence of a `t Hooft anomaly is an obstruction to gauging a symmetry, but the path integral \eqref{defpartition} involves gauging the symmetry.
The main reason for this is the fact that the anomaly is usually proprortional to the curvature of the gauge field, and that the compensator integrals continue to restrict the gauge fields to be flat.
We will then discuss an interesting implication of this fact.

We take as an operational definition of a `t Hooft anomaly that
\begin{equation}
  Z_{0} [e,A] \neq Z_{0} [e, A + dg].
  \label{eqn:thooft-anom-definition}
\end{equation}
In a fuller analysis, one has to try adding to the free energy local integrals of the gauge fields (counterterms) and show that there is no choice that make it gauge-invariant; further, in many interesting cases, the anomaly is \emph{mixed}, meaning that there are multiple symmetries and different choices of counterterms can lead to \eqref{eqn:thooft-anom-definition} being the case for different symmetries --- in other words, the obstruction is only to gauging all the symmetries together.
We assume for this section that this analysis has been done, and we have for whatever reason chosen a counterterm that causes \eqref{eqn:thooft-anom-definition} to be the case.

In this case, the path integral \eqref{defpartition} seems physically nonsensical, since the volume of the gauge group in the denominator isn't cancelling out an integral over a gauge symmetry any more.
However, it is not necessarily mathematically nonsensical: we can still do the path integral over all the modes of the gauge field --- including the now non-trivial gauge mode integral --- and then divide by the volume of gauge transformations, and see if we get a finite answer.
If there is a choice of counterterm in the seed that makes the answer finite, the path integral is not nonsensical.

We will now argue that not only is the answer finite, but it is just \eqref{torusfinal}.
The first observation leading to this fact is that, even without gauge-fixing, the manipulations in section \ref{ssec:constraints} up to \eqref{eqn:delta-fn-final} hold independent of the functional dependence of $Z_{0}$ on the gauge fields.
The second observation we need is that `t Hooft anomalies usually vanish when the gauge field is flat,
\begin{equation}
  \log Z_{0} [A+dg] - \log Z[A] \propto F=dA = d*d\chi.
  \label{eqn:anomaly-curvature}
\end{equation}
This is a familiar fact, but we prove it in appendix \ref{app:anomalies} for completeness.

Now, expanding the gauge field as in \eqref{eqn:A-decomp}, we see that the delta function constraint sets the non-gauge mode $\chi$ to $0$.
Now, because of \eqref{eqn:anomaly-curvature}, the $g$-dependent part of the partition function (ignoring factors, which are similar to those in previous sections) becomes\footnote{We forget about the second gauge field for simplicity.}
\begin{equation}
  \frac{1}{\ell_{1}} \int \tilde{n}_{a} (f-e)^{a} \wedge dg = -\frac{1}{\ell_{1}} \int g \tilde{n}_{a} d (f-e)^{a} = 0.
  \label{eqn:kernel-g-dependence}
\end{equation}
This vanishes, because $de=0$ is also one of the constraints.
We repeat again that, because of \eqref{eqn:anomaly-curvature}, the seed doesn't depend on $g$ when $\chi$ vanishes --- even though it does depend on $g$ when $\chi$ doesn't.
Thus, we see that the action no longer depends on $g$, and so the $Dg$ integral does cancel the $\text{vol}(G)$ in the denominator.
The factors then work out as in the previous subsections, and we are left with the same final answer \eqref{torusfinal}.

Thus, we are led to the surprising conclusion that our proposal, despite involving gauging the symmetries whose currents drive the deformation, leads us to a sensible answer satisfying equations like \eqref{eqn:global-flow-text} even when the symmetry has a `t Hooft anomaly.
This is true as long as the `t Hooft anomaly is proportional to the curvature.

An interesting implication of this is the following: one could have alternately tried to define the path integral without any compensators like $\alpha$ by gauge-fixing, and that would have given us the wrong answer in the anomalous case!
In more detail, when we gauge-fix $\alpha=0$, the $\text{vol}(G)$ in the denominator cancels the $D\alpha$ in the numerator, leaving us only with a $DA$ integral and no denominator.
In the non-anomalous case, the seed doesn't depend on the gauge mode $g$ but the kernel does; we then find that the $g$ integral plays the role of imposing the flatness constraint --- in some sense, by choosing this gauge, we have absorbed the compensator into the gauge mode.
However, in the anomalous case, the seed will depend on $g$, and so this integral no longer imposes a flatness constraint.
On the other hand, in neither case does the seed depend on $\alpha$, and so the $\alpha$ integral always imposes flatness.
To be perfectly clear, this is not a contradiction: in the anomalous case, one shouldn't gauge-fix, and so the $\alpha=0$ path integral is just a different object from this one.

\section{The Deformed Spectrum} \label{spectrum}

To complete this non-trivial check of our proposal, we perform the resulting finite dimensional integral.  We obtain the explicit form of the deformed spectrum, for different values of the deformation parameters and charges. The energy levels so obtained precisely match those in the literature, found by very different methods. We also briefly discuss the saddle point approximation to these integrals and their one-loop exactness.

We will restrict our attention to the case when the seed is a CFT, for the simple reason that in this case the seed partition function can be written as a simple sum.
Per the considerations in section \ref{quantum}, the path integral reduces to an integral over constant gauge fields, which are completely characterised by holonomies, so that the coupling term in the action becomes $\int A_{\mu} J^{\mu} \sim h_{0} \int J^{0} + h_{1} \int J^{1}$.
In general QFTs, there is no reason for the operators $\int J^{0}, \int J^{1}$ to commute, and therefore they can't be simultaneously diagonalised; this means that the path integral with non-zero holonomies cannot be written as a sum of the form $\sum e^{- \beta E - \mu Q}$.
In 2D CFTs, however, there is a holomorphic factorisation that allows one to get around this problem: the holomorphic and anti-holomorphic components $J,\bar{J}$ are separately conserved, at least in the absence of a background gauge field.
Because of this, both charges can separately take definite values, and we can write the partition function as a sum.
This gives us a good starting point to find the deformed spectrum, and so we will stick to this case.

This does however create a new question of its own.
Since the CFT has two completely different conserved currents ---  the chiral current $j_{c} = (J,0)$ and the vector current $j_{v} = (J, \bar{J})$ --- we are now presented with a choice of which current participates in the deforming operator $J \bar{T}$.
Keeping this ambiguity in mind, we have chosen to calculate the deformed spectrum in three cases:
\begin{enumerate}
  \item The first case is a deformation solely by $j_{c} \bar{T}$, dealt with in section \ref{ssec:jctbar}.
    Here, $j_{c}$ is the chiral current $(J,0)$.
    Further, in section \ref{ssec:jctbar-tjcbar-ttbar} we consider a joint flow with both $j_{c} \bar{T}$ and  $T \bar{T}$ deformations.

    An important subtlety with this case is that the chiral current has a `t Hooft anomaly, as explained in appendix \ref{app:anomalies}.
    This would mean that the path integral we have defined does not really have a gauge-invariance and so it'd seem that out analysis must fail for this case.
    However, somehwat surprisingly, this is not the case; we have explained in section \ref{ssec:anom} that the whole procedure works despite the existence of a `t Hooft anomaly.

  \item The second, dealt with in section \ref{ssec:jctbar-tjcbar} is a deformation by both $j_{c} \bar{T}$ and $T \bar{j}_{c}$ with two independent parameters $\ell_{1}, \ell_{2}$. Here $\bar{j}_{c}$ denotes a chiral current with only anti-hololomorphic component, that is $\bar{j}_{c}=(0,\bar J)$.

    Again, the sensibility of this calculation is a non-trivial fact explained in section \ref{ssec:anom}.

  \item Finally, we take the deforming operator to be $j_{v} \bar{T}$, which is like the first case but with a vector current, dealt with in appendix \ref{ssec:jvtbar}.
    We find that the deformation of the spectrum due to this a priori different deformation is but a special case $\ell_{1}  = \ell_{2}$ of the second case we considered,\footnote{We thank the JHEP referee for a useful question in this regard.}
    \begin{equation}
    E_{j_v \bar{T}} (\ell) = E_{j_c \bar{T}+ T \bar{j}_c} (\ell_1 = \ell_2 = \ell).
    \label{eqn:mysterious-equality}
    \end{equation}
    This equality is at this time mysterious, and an interesting avenue for future work.
\end{enumerate}
In all these cases, we reproduce the results in the literature.

 We will work in complex coordinates $(z,\bar z)$. Our conventions may be found in the Appendix. In terms of these complex variables, the integral we wish to compute reads
\begin{equation}
 \label{complexint}
Z= \frac{ {\cal A}2^3 }{(2 \pi)^4 \ell_1^2\ell_2^2}  \int d^2h d^2\tilde{h} \int \frac{d^2 \bar{L}d^2\tau'}{\tau'_2} \, e^{-\frac{2}{\ell_1}\big[(\bar\tau L_z-\bar\tau'\bar{L}_z) (b_{\bar z}-h_{\bar z})-(\tau L_{\bar z}-\tau'\bar{L}_{\bar z}) (b_z-h_z)\big]-\frac{2}{\ell_2}(L-\bar{L})\wedge (\tilde{b}-\tilde{h})}Z_0[\bar{L},h,\tilde{h}]
\end{equation}
Note all moduli dependence appears only in the first term of the kernel with our choice of parametrizations in \eqref{rot}.

\subsection{The $J\bar T$ Case with a Chiral Current} \label{ssec:jctbar}
In the language of appendix \ref{app:anomalies}, this is the case $\bar{k}=0$; so, with the addition of the counterterm \eqref{eqn:standard-shift} there is still a `t Hooft anomaly proportional to $k$, but because of the considerations of section \ref{ssec:anom}, we may still plug it into the path integral and obtain \eqref{complexint}.

To explicitly evaluate \eqref{complexint}, we need to address the definition of $Z_{0}$, the seed partition function. First off, we restrict our attention to seed CFTs, and hope to explore the more general QFT setting in future work. Secondly, we focus on the case where the undeformed theory has a single $U(1)$ current coupled to $h_\mu$.  Following \cite{Benjamin:2016fhe}, \cite{Kraus:2006wn}, we take the partition function $Z_0$ as the one defined via a path integral with appropriate counterterms.\footnote{This requires some care. First, even in the pure $T\bar{T}$ case, there existed a choice in defining the seed partition function. By writing the seed torus partition function as in \eqref{gentorus}, the authors of \cite{Dubovsky:2018bmo} chose a particular re-normalization scheme and subtracted off a possible cosmological constant term, c.f. Eq. 21 of \cite{PolchinskiTorus}. Similarly, we are choosing local counterterms in our path integral so that it results in what is quoted in the main text. The main distinction we were trying to draw with this comments was to contrast our seed definition with a more Hamiltonian definition, as discussed in section 5 of \cite{Kraus:2006wn} } Its dependence on $h_{\bar z}$ is essentially fixed by modular invariance. More precisely, adopting the conventions of \cite{Benjamin:2016fhe}, we take 
\begin{align}
Z_0&=e^{-2\pi k \tau'_2 h_{\bar z}^2}\sum_n e^{ 2\pi i\tau' \bar R E_{0,n}^{(L)}(\bar R)- 2\pi i\bar\tau' \bar R E_{0,n}^{(R)}(\bar R)+ 2\pi \tau'_2 h_{\bar z} Q_n }\\
&=e^{-2\pi k \tau'_2 h_{\bar z}^2}\sum_n e^{ 2\pi i \tau'  \varepsilon_{0,n}^L- 2\pi i \bar\tau' \varepsilon_{0,n}^{(R)}+ 2\pi \tau'_2 h_{\bar z} Q_n }
\label{undef}
\end{align}
where $\bar{R}^2=4\bar{L}_z\bar{L}_{\bar z}$ is the base space radius. In going to the second line, we have introduced the dimensionless energies. They are $\bar R$-independent since the seed theory is a CFT.  $k$ denotes the Kac-Moody level of the chiral algebra satisfied by the current.
Note that the prefactor proportional to $h_{\bar z}^2$ comes not from an anomaly but by the requirement of modular invariance, as proved in \cite{Benjamin:2016fhe}.

Note the holonomies $h_z,\tilde h_z,\tilde h_{\bar z}$ act as Lagrange multipliers, enforcing the constraints
\begin{align}
\bar L_z &= L_z \\
\bar L_{\bar z}&= L_{\bar z}\\
\bar L_{\bar z}\tau'&= L_{\bar z} \tau
\end{align}

This would not have been the case had the holonomies been restricted to the compact space $U(1)$; see the discussion in section \ref{gauge-unwrap}. Furthermore, the delta-functions imposing the above constraints come with the following prefactor
\be
\frac{(2\pi)^3\ell_1\ell_2^2}{8 L_{\bar z}}
\ee
For simplicity, let us consider the target space configuration in wich $L_z=i \frac{R}{2} = -L_{\bar z}$  ({\it i.e.} $L_2=0$) along with vanishing fluxes $b_\mu=\tilde b_\mu =0$. We also choose to solve the third constraint for the $\tau'_1$ variable, that is
\be
\bar L_z=-\bar L_{\bar z}= i \frac{R}{2}
\ee
\be
\tau'_1= \tau_1 +i(\tau_2-\tau'_2) 
\ee
Putting it all together, we are left with the following integral 
\begin{align} 
Z&= \sum_{n} \frac{{\cal A} }{ \pi \ell_1 R} \int \frac{dh_{\bar z}d\tau'_2}{\tau'_2} e^{\frac{2R}{\ell_1}(\tau_2-\tau'_2) h_{\bar z}+2\pi \tau'_2 Q_n h_{\bar z} -2\pi k\tau'_2 h_{\bar z}^2-4\pi \tau'_2 \varepsilon_{0,n}^{(R)}+2\pi i k_n(\tau_1+i\tau_2)} \label{htauint} \\
&=\sum_{n} \frac{{\cal A} }{\sqrt{2 k} \pi \ell_1 R} \int \frac{d\tau'_2}{(\tau'_2)^{3/2}}e^{\frac{(R(\tau_2-\tau'_2)+\ell_1\pi Q \tau'_2)}{2\pi k \ell_1^2 \tau'_2}}e^{-4\pi \tau'_2\varepsilon_{0,n}^{(R)}+2\pi i k_n(\tau_1+i\tau_2) }\\
&=\sum_{n} e^{2\pi i k_n \tau_1 -2 \pi \tau_2 (2R E_{n}^{(R)}+k_n)} \label{Zexact}
\end{align}
where we have used ${\cal A}=R^2\tau_2$ in order to cancel the prefator of the exponential. The main outcome of this computation is the right-moving deformed  energy $E_{n}^{(R)}$ which reads
\be
E_{n}^{(R)} = \frac{1}{4\pi^2 k \ell_1^2}\left(R-\ell_1\pi Q_n-\sqrt{(R-\ell_1\pi Q_n)^2-8\pi^2 k\ell_1^2\varepsilon_{0,n}^{(R)} }\right)
\label{ER}
\ee
The square root branch has been chosen such that it satisfies the correct innitial condition.
The above spectrum satisfies 
\be
R E_{n}^{(R)} -\pi \ell_1 Q_n E_{n}^{(R)} -2\pi^2 k(\ell_1 E_{n}^{(R)})^2 = \varepsilon_{0,n}^{(R)} = (\ell_1 \, \, {\rm independent}) \label{EReq}
\ee
which is the analog (in our conventions) of equation (6.20) in \cite{Chakraborty:2018vja}. It is satisfying to see the spectrum obtained here matches the one previously reported in the literature. 

For future reference, let us point out the $k=0$ case of this equation can readily be solved. The $h,\tilde h$ dependence is linear. Integration leads to a set of four constraints. Their solution localizes the base space modular parameters to the following (state-dependent) locus
\be
\tau'_1 = \tau_1 -i \frac{\ell_1 \pi Q_n}{R-\ell_1 \pi Q_n}\tau_2 \quad , \quad \tau'_2=  \frac{R}{R-\ell_1 \pi Q_n}\tau_2 
\label{k0tau}
\ee  
which, once plugged into the partition function, leads to the simple spectrum found in \cite{Guica:2017lia}
\be
E_{n}^{(R)}\Big|_{k=0} = \frac{\varepsilon_{0,n}^{(R)}}{R-\ell_1\pi Q_n} \,.
\label{k0E}
\ee
The last equation introduces the notion of an effective (state-dependent) radius given by $R_{eff}=R-\ell_1 \pi Q$, previously described in \cite{Guica:2017lia}. A similar structure arises for $k\neq0$, as explained further below.

\subsection{One-loop Exactness and the Chiral Charge}

 \cite{Dubovsky:2018bmo} found the integral giving rise to the $T\bar T$ deformed torus partition function to be one-loop exact. That implies, in particular, the exact spectrum can be consistently obtained from its saddle point approximation. 

We find that our \eqref{htauint} shares this remarkable one-loop exactness property. To be clear, this means it is completely dominated by its saddle point value and the determinant for quadratic fluctuations around the saddle. It thus behaves like a Gaussian integral, even though it certainly is not. This is yet another diagnostic of the flow's solubility, as seen from our path integral perspective. 
The saddle point approximation to the integral clarifies some important conceptual puzzles and makes contact with previous discussions in the literature. In particular, it immediately singles out a set of equations which, within a completely different approach, led to spectrum just found \eqref{ER}. 

In order to proceed, let us take \eqref{htauint} which is admittedly Gaussian in $h_{\bar z}$ but not in $\tau_{2}'$. For sake of the argument, we include a putative "$1/ \hbar$" parameter in the action, which we later set back to 1: 
\be
Z=\sum_{n} \frac{{\cal A} }{ \pi \ell_1 R}e^{2\pi i k_n(\tau_1+i\tau_2)} \int \frac{dh_{\bar z}d\tau'_2}{\tau'_2} e^{\frac{1}{\hbar}\Big[\frac{2R}{\ell_1}(\tau_2-\tau'_2) h_{\bar z}+2\pi \tau'_2 Q_n h_{\bar z} -2\pi k\tau'_2 h_{\bar z}^2-4\pi \tau'_2 \varepsilon_{0,n}^{(R)}\Big]}
\label{Zsaddle}
\ee
We compute the saddle point equations for $\tau'_2$ and $h_{\bar z}$ respectively
\begin{align}
-\frac{R}{\ell_1} h_{\bar z}+\pi Q_n h_{\bar z} -\pi k h_{\bar z}^2 - 2\pi \varepsilon_{0,n}^{(R)} &=0
\label{tausaddle}\\
\frac{R}{\ell_1}(\tau_2-\tau'_2) + \pi \tau'_2 (Q_n-2k h_{\bar z}) &=0 
\label{hsaddle}
\end{align}

It remains to evaluate the fluctuation determinant
\begin{align}
\left(\frac{2\pi}{\hbar}\right)\det\left(\partial^2 S\right)^{-1/2}\Big|_{\text{on-shell}}&=\left(\frac{2\pi}{\hbar}\right) \left(\frac{2R}{\ell_1}-2\pi Q_n +4\pi k h_{\bar z}\right)^{-1}\Big|_{\text{on-shell}} \nn \\ 
&= \frac{\pi\ell_1 \tau'_2}{ R \tau_2}
\end{align}
thus canceling the prefactor in \eqref{Zsaddle}. The one-loop approximation thus recovers the exact expression in \eqref{Zexact}. The semiclassical spectrum is actually the exact one \eqref{ER}. We may gain intuition by identifying
\be
h_{\bar z}|_{\text{saddle}} = -2\pi \ell_1 E_{n}^{(R)}
\ee
By further making the following definition,  
\be
{\cal Q}_n = Q_n+4\pi k \ell_1 E_{n}^{(R)} \label{Qeffdef}
\ee
the second saddle point equation \eqref{hsaddle} takes on a very suggestive form
\be
\tau'_2 = \frac{R}{R_{eff}}\tau_2 \quad , \quad R_{eff}=R-\ell_1 \pi {\cal Q}_n \label{Reff}
\ee
which is to be compared with the $k=0$ solution for $\tau'_2$ \eqref{k0tau}. Combining \eqref{Qeffdef} and \eqref{tausaddle} we also get
\be
R E_{n}^{(R)} -\frac{1}{8k}{\cal Q}_n^2 =\varepsilon_{0,n}^{(R)} -\frac{1}{8k}Q_n^2= \, constant \,. 
\label{EQrel}
\ee 
Requiring all quantities depend on the dimensionless ratio $\ell_1/R$, we can recast the above expressions as two differential equations\footnote{Note all the relations obatined here completely match the ones listed in \cite{Chakraborty:2018vja} by taking $\mu^{\rm there}=2\pi \ell_1$ and $k=1/4$.  }
\begin{align}
\partial_{\ell_1} E_{n}^{(R)} &= -\pi {\cal  Q}_n \partial_R E_{n}^{(R)} \label{diff1}\\
{\cal  Q}_n \partial_{\ell_1}{\cal  Q}_n &= 4 k R\partial_{\ell_1} E_{n}^{(R)}  \label{diff2}
\end{align}
We thus recover the defining equations for the energy levels. These differential equations were solved in \cite{Bzowski:2018pcy,Chakraborty:2018vja}. The solution for the energy is none other than \eqref{ER} (consistent with one-loop exactness found above), together with 
\be
{\cal Q}_n = \frac{1}{\pi \ell_1}\left(R-\sqrt{(R-\ell_1\pi Q_n)^2-8\pi^2 k\ell_1^2\varepsilon_{0,n}^{(R)} }\right) \,.\label{Qeff} 
\ee
In order to better understand the role of ${\cal Q}_n$, let us briefly comment further on equations \eqref{diff1} and \eqref{diff2}. 
In the $T\bar T$ scenario, solving a single differential equation leads to the dressed energy levels. In  \cite{Zamolodchikov:2004ce,Smirnov:2016lqw}, that equation results from the factorization property of the deforming operator. Along with rotational invariance, the flow can be written purely in terms of the energy and momentum of a given state. Without rotational invariance, additional assumptions would be needed \cite{Cardy:2018jho}.    

Even though the $J\bar T$ operator still nicely factorizes, there is no way of getting a sensible differential equation for the energy levels without impossing an additional constraint: the deforming current needs to be chiral. Otherwise, the flow equation involves the expectation value of its spatial component, which usually is not quantized. Requiring chirality allowed \cite{Guica:2017lia} to solve for the spectrum, which we recovered here  for $k=0$ in \eqref{k0E}. 
 
Requiring the flow to preserve chirality is in many ways too strong a constraint on the set of possible trajectories. In the more general case, defining the charge ${\cal Q}_n$ associated with the chiral projection of the deforming current circumvents this issue \cite{Bzowski:2018pcy,Chakraborty:2018vja}. This identification in some sense ``emulates'' the chiral case. It is therefore not surprising the effective radius $R_{eff}$ \eqref{Reff} depends on ${\cal Q}_n$ in the same way as for $k=0$.  The price one has to pay is having to solve two differential equations instead of one, namely \eqref{diff1}, \eqref{diff2}. These equations arise naturally from our approach. 

The role of $k$ along the flow thus becomes clearer. Physically, when $k\neq 0$, the chiral current develops an anti-holomorphic part, 
spoiling chirality at the quantum level. As $k\to 0$, the second equation \eqref{diff2} becomes trivial, and we recover the case studied in \cite{Guica:2017lia}. In our path integral approach, this phenomenon is manifested by the $h^2$ term in the seed partition function. Without it, the integration over holonomies would lead to a simple constraint over the geometric variables. It would not appear in an additional saddle point equation.

Finally, here we show a different approach within these quantities arise naturally. So far, we have been focused on the spectrum of the deformed theory. The background fluxes $b,\tilde b$ played no role in the previous discussion. For sake of completeness, we write down the result of the finite dimensional integral in presence of non-trivial target space holonomies. Again, we consider $\tilde Q_n=0$, so the $\tilde b$ fluxes still play no role. They can be absorbed in the $\tilde h$ integration variable. In fact, the same applies to the holomorphic component of $b$. We thus only need consider non-trivial anti-holomorphic $b_{\bar z}$.  
The intermediate steps being rather unenlightening, we simply report the result
\be
Z_b = \sum_{n} e^{2\pi i \tau_1 k_n -2\pi \tau_2 (2 R F_{b,n}+k_n)}
\ee  
with
\be
F_{b,n} = \frac{1}{4\pi^2 k \ell_1^2}\left( R-\ell_1 \pi(Q-2k b_{\bar z})-\sqrt{(R-\ell_1 \pi(Q-2k b_{\bar z}))^2-4\pi^2 k\ell_1^2(2\varepsilon_{0,n}^{(R)}+b_{\bar z}(b_{\bar z}k-Q))}\right)
\ee
The deformed theory is not conformally invariant. We are thus not able to fix the dependence on the holonomies for the torus partition function. \eqref{undef} no longer serves as point of comparison. Therefore, the actual physical meaning of the function $F_{b,n}$ multiplying $\tau_2$ is unclear. In particular, it cannot be identified with an energy level.  

However, some physical intuition can be gained  by expanding the result for small values of $b_{\bar z}$. This gives
\be
2 R F_{b,n} = 2 R E_{n}^{(R)} -\left(\frac{R}{R_{eff}}\right){\cal Q}_n \,  b_{\bar z} + \left(\frac{R}{R_{eff}}\right)^3 k b_{\bar{z}}^2  + {\cal O}(b_{\bar z}^3)
\ee
with $E_{n}^{(R)}$, $R_{eff}$ and ${\cal Q}_n$ given by \eqref{ER}, \eqref{Reff} and \eqref{Qeff} respectively.

\subsection{General $J\bar T+T\bar J$ Deformation} \label{ssec:jctbar-tjcbar}

Consider finally the general case with non-trivial charged states for both $J$ and $\tilde{J}$. Denote these charges $Q_n$ and $\tilde{Q}_n$ respectively. As this computation closely parallels that of the above section, we will be brief. 
The seed partition function becomes  
\be
Z_0= e^{-2\pi\tau_2' k(h_{\bar z}-\tilde{h}_z)^2} \sum_{n} e^{2\pi i\tau' \varepsilon_{0,n}^{(L)}-2\pi i \bar\tau' \varepsilon_{0,n}^{(R)} +2\pi\tau'_2 Q_n h_{\bar z}+2\pi\tau'_2 \tilde{Q}_n \tilde h_z}
\ee
 Our choice of parametrization here will be different from \eqref{rot}, namely\footnote{Note we are not including the $\frac{1}{\sqrt2}$ factors here, as they are again absorbed by rescaling the couplings.}
\begin{align}
\tilde{n}_a L^a_z=i \bar\tau L_z \quad &, \quad \tilde{n}_a L^a_{\bar z}= iL_{\bar z}\\
n_a L^a_z= i L_z \quad &, \quad n_a L^a_{\bar z}= i \tau L_{\bar z}
\end{align}
Our kernel's action (again with $b=\tilde b=0$) now reads
\be
\frac{2h_{\bar z}}{\ell_1}(\bar\tau L_z-\bar\tau' \bar L_z)-\frac{2h_{z}}{\ell_1}( L_{\bar z}- \bar L_{\bar z}) +\frac{2\tilde{h}_{\bar z}}{\ell_2}( L_z-\bar L_z)-\frac{2\tilde{h}_{z}}{\ell_2}(\tau L_{\bar z}-\tau' \bar L_{\bar z}) 
\ee
To do the integrals, it is convenient to shift $h_{\bar z}\to h_{\bar z}+\tilde{h}_z$. The $h_z,\tilde{h}_z,\tilde{h}_{\bar z}$ integrations now impose the following constraints
\begin{align}
\bar L_{z}&=L_{z}\\
\bar L_{\bar z}&=L_{\bar z}\\
\frac{2}{\ell_1}(\bar\tau L_{z}-\bar\tau'\bar{L}_{ z})-&\frac{2}{\ell_2}(\tau L_{\bar z}-\tau'\bar{L}_{\bar z})=-2\pi\tau'_2 (Q_n+\tilde Q_n) 
\end{align}
which have solutions (for $L_z=-L_{\bar z}=i\frac{R}{2}$)
\begin{align}
\bar L_{z}&=-\bar L_{\bar z}=i \frac{R}{2}\\
\tau' &= \tau-2i\frac{\ell_2}{\ell_1+\ell_2}\tau_2+2 i\ell_2\frac{R-\ell_1 \pi (Q_n+\tilde Q_n)}{R(\ell_1+\ell_2)}\tau'_2
\end{align}
The prefactor arising from the delta-functions imposing the above contraints is 
\be
\frac{(2\pi)^3 \ell_1^2\ell_2^2}{4R(\ell_1+\ell_2)}
\ee
Finally, performing the remaining $h_{\bar z}$ and $\tau'_2$ integrals, we arrive at  
\be
Z =\sum e^{2\pi i \tau_1 k_n -2\pi \tau_2 RE_{n}}
\ee
with the deformed energy quoted early on \eqref{doubledef}, which reads
\begin{align}
E_{n}= &\frac{1 }{2\pi^2k(\ell_1+\ell_2)^2}\Big(R-\ell_1 \pi Q_n +\ell_2\pi\tilde{Q}_n+ 2\pi^2 k (\ell_1+\ell_2)(\ell_1-\ell_2)\frac{k_n}{R}\,+ \label{eqn:jtbar-tjbar-energies}\\
&\left.\sqrt{(R-\ell_1 \pi Q_n+\ell_2 \pi \tilde{Q}_n)^2-8\pi^2 k (\ell_1+\ell_2)^2\left(\varepsilon_{0,n}^{(R)}+k_n\ell_2\frac{R-\ell_1\pi(Q_n+\tilde Q_n)}{R(\ell_1+\ell_2)}\right)}\right)\nn
\end{align}
The branch of the square root has again been chosen so the deformed energy satisfy the correct initial conditions. The above spectrum precisely matches the one found in \cite{Chakraborty:2019mdf,Hashimoto:2019wct}.  \footnote{In order to check the agreement with the results listed in \cite{Chakraborty:2019mdf} one should take $q_R=-\tilde{Q}_n$, $k=1/4$ together with the following relation between parameters 
$$
\hat{\epsilon}_+=\frac{\pi\ell_1}{R} \quad , \quad \hat{\epsilon}_-=\frac{\pi\ell_2}{R} \quad , \quad \hat\lambda = -\frac{8\pi \lambda}{R^2}
$$}

\subsection{Joint flow with $T\bar T$ } \label{ssec:jctbar-tjcbar-ttbar}
 
 Studying the joint flow of $J \bar{T}$ and $T\bar T$ provides our last application of the path integral representation of the deformed partition function. This amounts to taking $\gamma\neq 0$ in the kernel. We consider the general $k$ and $\tilde Q_n=0$ case. Using the parametrization in \eqref{rot}, almost identical manipulations as those described so far lead to the kernel
\be
\frac{2h_{\bar z}}{\ell_1}(\bar\tau L_z-\bar\tau'\bar L_z)+\frac{2h_z}{\ell_1}(\tau L_{\bar z}-\tau'\bar L_{\bar z})+\frac{2\tilde{h}_{\bar z}}{\ell_2} (L_z-\bar L_z)-\frac{2\tilde h_z}{\ell_2}(L_{\bar z}-\bar L_{\bar z})-2i\gamma (h_z \tilde h_{\bar z}-h_{\bar z}\tilde h_z)\,.
\ee
Integrating over $\tilde h_z$ and $\tilde h_{\bar z}$ gives rise to two delta functions. They allow us to immediately perform the $h_z$ and $h_{\bar z}$ integrals which set
\be
 h_z=i\frac{\bar L_z-L_z}{\gamma \ell_2}\,, \quad  h_{\bar z}=i\frac{\bar L_{\bar z}-L_{\bar z}}{\gamma \ell_2}\,.
\ee
There is a factor $(2\pi)^2(2\gamma )^{-2}$ coming from the delta  functions.  
On the above solution locus, the kernel is still linear in $\bar L_z$. We can thus further integrate over this variable. It localizes the $\bar L_{\bar z}$ integral to
\be
\bar{L}_{\bar z} = L_{\bar z} \frac{\tau-\bar\tau'}{2i \tau'_2}
\ee 
along with a $(2\pi)\frac{ \gamma\ell_1\ell_2}{4 \tau'_2}$ prefactor.

By defining $\lambda= \gamma \ell_1 \ell_2$, and shifting $\tau'_1\to \tau_1+\tau'_1$, we arrive at 
\be
Z= \sum_{n} \frac{{\cal A}}{4\pi \lambda} \int \frac{d^2\tau'}{(\tau'_2)^2}e^{\frac{R(\tau_2+i\tau')(R(2\lambda(\tau_2-i \bar{\tau}')-\pi k\ell_1^2 (\tau_2+i\tau'))+4\pi Q_n \ell_1 \lambda \tau'_2)}{8\lambda^2 \tau'_2}+2\pi i k_n (\tau'_1+\tau_1)-2\pi \tau'_2 \varepsilon_{0,n}}
\ee
We can again perform this integral over base space torus moduli and extract the desired spectrum, giving
\begin{align}
E_{n}= \frac{1}{2\pi \tilde{\lambda}}\Big( & R-\pi\ell_1 Q_n - 2\pi^2 k \ell_1^2 P_{n} \\
& - \sqrt{(R-\pi\ell_1 Q_n)^2 - 2\pi \tilde\lambda(2\varepsilon_{0,n}-4\pi \lambda P_{n}^2)+4\pi^2\ell_1 P_{n}( k \ell_1 R+2\lambda(Q_n-\pi k \ell_1 P_{n}))}\Big)
\nn\end{align}
where we have reintroduced the momentum $P_{n}=k_n/R$ and defined the coupling
\be
\tilde \lambda = 2\lambda+\pi k \ell_1^2 \label{lambdaeff}
\ee
In particular, by taking the $\ell_1\to 0$ limit keeping $\lambda=\gamma \ell_{1} \ell_{2} $ fixed, the expression above reduces to the standard formula for the $T\bar T$ dressed energy levels
\be
\lim_{\ell_1\to 0} E_{n} = \frac{R}{4\pi \lambda}\left(1-\sqrt{1-\frac{8\pi\lambda \varepsilon_{0,n}}{R^{2}}+\frac{16\pi^2 \lambda^2}{R^2}P_{n}^2}\right)
\ee
displaying the usual Hagedorn behaviour for negative $\lambda$. Note the presence of such a behaviour is dictated by the sign of $\tilde\lambda$ given in \eqref{lambdaeff}. In particular, when both couplings are present and $\lambda<0$, we find a crossover point at $\ell_{1} =\sqrt{2|\lambda|/\pi k}$. The ability of $J \bar T$ deformation to ``remove'' the Hagedorn regime has been discussed in \cite{Chakraborty:2019mdf,Hashimoto:2019wct}.

\section{Conclusion and Future Directions}

This work presented a path integral realization of Lorentz breaking irrelevant $J\bar T$ and $T\bar J$ deformations. We have recast their joint flow with $T\bar{T}$  in \eqref{GenJTkernel} as coupling the seed to a topological quantum gravity and gauge theory. As it was for its pure $T\bar T$ predecessor, our path integral kernel fundamentally translates between a base space, where the undeformed theory lives, and a target space on which the deformed theory is defined. The path integral is an integral over a restricted set of maps between the spacetime and $U(1)$ frames of the base space - the vielbeins $e^a$ and gauge connections $A, \tilde{A}$ - to target space ones denoted by $f^a, B \ \text{and} \  \tilde{B}$. The compensator fields $Y^a$ and $\alpha, \tilde{\alpha}$ make diffeomorphism and gauge invariance of the kernel manifest. They further serve to implement important constraints on the path integral, which ultimately make it soluble.

Our proposal succesfully passed a wide variety of non-trivial checks. At the classical level, the kernel recovers the exact classical deformed actions. The procedure involves nothing more than solving an algebraic system of equations and reproduces the known expressions for the free boson and fermion.  At the quantum level, we reduced the full path integral over base space torus geometries and gauge connections to a finite dimensional one. It solves the desired difussion-like equation. Most importantly, by explicitly evaluating the torus partition function for certain seed theories, we extracted the deformed spectrum  along particular flows triggered by combinations of $J\bar T$, $T\bar J$ and $T\bar T$. Our results all matched the known expressions in the literature.

In many ways, this path integral construction puts the $J\bar T$ and $T\bar J$  deformations on similar footing to $T\bar T$'s . However, it also brings with it a plethora of new questions. First off, we saw a proper treatment of target space diffeomorphism invariance required $df^a=dB=d\tilde{B}=0$. Ultimately, we hope to engineer a kernel that generalizes away from flat vielbeins and vanishing $U(1)$ field strengths. In \cite{Mazenc:2019cfg}, we will report on progress on defining \textit{a} $T\bar{T}$ deformation for curved spacetimes with $df^a \neq 0 $. This encourages us to seek, in future work, a kernel that also accommodates $dB,d\tilde{B} \neq 0$. Furthermore, the one-loop exactness properties first found in \cite{Dubovsky:2017cnj} led the authors to conjecture the $T\bar{T}$ deformation of a general QFT might be captured as coupling to a form of 2d gravity. The one-loop exactness discovered for our kernel gives us hopes to similarly extend our formalism beyond CFT seeds. However, the expression of the seed partition function as a sum appears more involved in this case.

Finally, another interesting question raised by our results is the reason for the equality \eqref{eqn:mysterious-equality}, which expresses the fact that the energy spectrum induced from two different deformations --- namely the $J \bar{T}$ deformation with $J$ being a vector current and the $J \bar{T} + T \bar{ \tilde{J} }$ with $J,\bar{\tilde{J}}$ being chiral currents --- are precisely the same.

Finally, \cite{Gorbenko:2018oov} engineered a modified $T\bar T$ flow (adding the "$\Lambda_{2}$ flow") to derive a holographic field theory dual to de-Sitter bulk geometries. We are quite curious what our more general flows, including $J\bar T$ and $T\bar J$, might have to say about these dS/dS holographic constructions and their novel bulk reconstruction features \cite{Lewkowycz:2019xse}.  

We close by admitting the partition function captures only a small portion of the physics lurking in the deformed theory. One of the most pressing issues to address is the structure of correlation functions. First steps have been taken in \cite{Cardy:2019qao,Guica:2019vnb}. We are exploring the introduction of additional sources in our path integral kernel to open up another front of attack on this important problem.

\subsection*{Acknowledgments}

It is a pleasure to acknowledge very helpful conversations with Victor Gorbenko, Jorrit Kruthoff, Eva Silverstein and Gonzalo Torroba.  EAM and RMS especially thank Vasudev Shyam for collaboration on the related project \cite{Mazenc:2019cfg}; much wisdom has flowed between these two projects. JAD and RMS would also like to thank the Aspen Center for Theoretical Physics for hospitality, where part of this work was completed.  JAD and ISL would also like to thank the Abdus Salam International Center for Theoretical Physics for hospitality. ISL would like to thank Balseiro Institute for hospitality during several stages of this work. This work was partially performed at the Aspen Center for Physics, which is supported by National Science Foundation grant PHY-1607611.

We also thank JHEP referees for urging us to think more carefully about `t Hooft anomalies.

\appendix
\section{Conventions} \label{Appendix}

Working in complex coordinates, it is worth keeping track of the different factors arising from this change of coordinates. On the one hand, as the integral over the base space manifold has been already done in \eqref{torusfinal}, the Levi-Civita tensor that appears is no longer a density , thus taking the following form
\be
\epsilon^{z\bar z} = -\epsilon^{\bar z z}= -2i\,.
\ee
Furthermore, the torus lengths and holonomies are
\begin{align}
L^{1}_z = \frac12(iL^{1}_1+L^{1}_2) \quad &, \quad L^{1}_{\bar z} = \frac12(-iL^{1}_1+L^{1}_2)\\
\Rightarrow L_z^2 = \bar \tau L_z^1 \quad &, \quad L_{\bar z}^2 = \tau L_{\bar z}^1\\
h_z = \frac12(ih_1+h_2) \quad &, \quad h_{\bar z} = \frac12(-ih_1+h_2)
\end{align}
and similarly for $\bar L$ and $\tilde h$. So the integration measure becomes
\be
d^2 \bar L d^2h d^2\tilde h  = 2^3 d\bar L_z d\bar L_{\bar z}dh_z dh_{\bar z}d\tilde h_z d\tilde h_{\bar z}\,.
\ee
Finally, we make the following choice for tangent space indices
\be
n_a L^a_\mu = \frac{i}{\sqrt{2}} L^1_\mu \quad , \quad \tilde{n}_a L^a = \frac{i}{\sqrt{2}} L^2_\mu \,. \label{rot}
\ee
This is a choice of tangent space orientation; the tangent space metric is whatever it needs to be for these two vectors to be light-like.

For convenience, we also redefined our couplings in section \ref{spectrum}
\be
\ell_i\to \frac{\ell_i}{\sqrt{2}} \,.
\ee

\section{`t Hooft Anomalies} \label{app:anomalies}
In this appendix, we prove that there is always a choice of counterterm for a CFT that makes the `t Hooft anomaly proportional to the curvature of the background gauge field.
We will use only the form of the current algebra and conservation in the absence of background gauge fields.

In a CFT, the two components of the current have the OPEs
\begin{align}
  J(z) J(w) &\sim \frac{k}{2(z-w)^2} \nonumber\\
  \bar{J}(z) \bar{J}(w) &\sim \frac{\bar{k}}{2(\bar{z}-\bar{w})^2}.
  \label{eqn:jj-ope}
\end{align}
Generically, $k, \bar{k}$ are two independent numbers.
Further, because of holomorphic factorisation in CFTs, the conservation equations (in the absence of a background gauge field) are simply
\begin{equation}
  \partial \bar{J} = \bar{\partial} J = 0.
  \label{eqn:hol-cons}
\end{equation}
This is the normalisation with which
\begin{equation}
  \delta \log Z[a] = \int \langle J \rangle \bar{a} + \langle \bar{J} \rangle a.
  \label{eqn:j-def}
\end{equation}
Under a gauge transformation, the partition function transforms to first order as
\begin{align}
  \delta_{g} \log Z[a] &= \int \langle J \rangle_{a} \bar{\partial} g + \langle \bar{J} \rangle_{a} \partial g \nonumber\\
  &= - \int g \langle \bar{\partial} J + \partial \bar{J} \rangle_{a}.
  \label{eqn:logz-g-tr}
\end{align}
Here, the $a$ subscripts mean that the expectation values need to be evaluated in the presence of the background gauge field $a$.
Let us take $a=a_{0}$ to be a background in which this current is conserved; by assumption $a_{0} = 0$ is such a choice.
Now, let us see the effect of a gauge transformation in the presence of the deformed background gauge field $a_{0} + \delta a$.
\begin{align}
  \delta_{g} \delta_{a} \log Z[a] &= - \int d^{2}z d^{2}w g(z)\left\langle (\bar{\partial} J + \partial \bar{J})(z) (\delta\bar{a} J + \delta a \bar{J})(w) \right\rangle \nonumber\\
  &= - \int d^{2}z d^{2}w g(z) \left\{ \delta \bar{a}(w) \bar{\partial}_{z} \frac{k}{2(z-w)^{2}} + \delta a (w) \partial_{z} \frac{\bar{k}}{2 (\bar{z}-\bar{w})^{2}} \right\}
  \label{eqn:logz-g-tr-2}
\end{align}
Since this is a derivative with respect to $\bar{z}$ acting on a function of $z$ and vice versa, one might think that this vanishes.
This is however only true up to a contact term, since $z^{-2} = - \partial_{z}^{2} \log |z|^{2}$ and $\partial_{z} \bar{\partial}_{z} \log |z|^{2} = 4\pi \delta^{2} (z)$.
Plugging these facts into the expression \eqref{eqn:logz-g-tr-2}, we find
\begin{align}
  \delta_{g} \delta_{a} \log Z &= 2\pi \int d^{2}z d^{2}w g(z) \left( k \delta \bar{a}(w) \partial_{z} + \bar{k} \delta a(w) \partial_{\bar{z}} \right) \delta^{2}(z-w) \nonumber\\
  &= - 2\pi \int d^{2}z g(z) \left( k \partial \delta \bar{a}(z) + \bar{k} \bar{\partial} \delta a(z) \right).
  \label{eqn:thooft-anom}
\end{align}
This is the famous `t Hooft anomaly.

Note that we are allowed to add to the partition function any local functional of the background gauge field.
In particular, by taking
\begin{equation}
  \log Z[a] \to \log Z[a] -2\pi \frac{k + \bar{k}}{2} \int a \bar{a},
  \label{eqn:standard-shift}
\end{equation}
we shift the currents as
\begin{equation}
  J \to J -2\pi \frac{k + \bar{k}}{2} a, \quad \bar{J} \to \bar{J} -2\pi \frac{k + \bar{k}}{2} \bar{a},
  \label{eqn:standard-shift-current}
\end{equation}
and then \eqref{eqn:thooft-anom} becomes
\begin{equation}
  \delta_{g} \delta_{a} \log Z = -2\pi \frac{k- \bar{k}}{2} \int d^{2} z g(z) \left( \partial \delta \bar{a} - \bar{\partial} \delta a \right) \propto F_{z \bar{z}}.
  \label{eqn:thooft-anom-corrected}
\end{equation}
Thus, we see that there is always a choice of counterterm in which the `t Hooft anomaly is proportional to the curvature.

\section{The $J \bar{T}$ Case with a Vector Current} \label{ssec:jvtbar}
In the language of appendix \ref{app:anomalies}, this is the case $k=\bar{k}$; so, with the addition of the counterterm \eqref{eqn:standard-shift} there is no `t Hooft anomaly.
This is a somewhat novel result that arises out of our approach; while previous literature stuck to using chiral currents in the $J \bar{T}$ deformation, we can as easily solve the $J \bar{T}$ deformation when the current $J$ is a vector current.

Now the seed partition sum coupled to the holonomies reads
\be
Z_0=e^{-2\pi k \tau'_2 (h_{\bar z}-h_z)^2}\sum_n e^{ 2\pi i \tau'  \varepsilon_{0,n}^L- 2\pi i \bar\tau' \varepsilon_{0,n}^{(R)}+ 2\pi \tau'_2 h_{\bar z} Q_n+2\pi \tau'_2 h_{z} \bar Q_n }
\ee
where $\bar Q_n$ denotes the charges associated to the anti-holomorphic part of the current which, when coupled to flat gauge connections, remains conserved ({\it c.f.} discussion presented in section \ref{ssec:anom}). The presence of two conserved charges associated to components of a single vector current $(J,\bar J)$ is nothing we are not familiarized within the context of CFT, as is the case of momentum and winding in a free compact scalar theory.

The computation can be performed along the same lines depicted in section \ref{ssec:jctbar}. Again, we choose to align the torus parameters as in \eqref{complexint}, together with $L_z=-L_{\bar z}=i R/2$. Additionally, we shift $h_{\bar z}\to h_{\bar z}+h_z$ in order to be left with only one Gaussian integral over the holonomies. Therefore, the constraints from integration over $\tilde h_z$, $\tilde h_{\bar z}$ and $h_z$ are solved by
\begin{align}
\bar L_z&=-\bar L_{\bar z}=i R/2\\
\tau_1' &= \tau_1-i\pi \ell_1 \tau_2' \frac{Q+\bar Q}{R}
\end{align}  
with the corresponding $\pi^3 \ell_1\ell_2^2 R^{-1}$ prefactor. 

Finally we have
\begin{align}
Z &= \frac{{\cal A}}{2\pi R \ell_1 } \sum_n \int \frac{dh_{\bar z}d\tau_2'}{\tau_2'}
e^{ -2\pi k \tau_2' h_{\bar z}^2 +\frac{R(\tau_2-\tau_2')+\pi\ell_1(Q_n-\bar Q_n)\tau_2'}{\ell_1} + 2\pi i k_n \tau_1-2\pi\tau_2' \varepsilon^0_n + 2\pi \tau_2' k_n \ell_1\frac{Q_n+\bar Q_n}{R}}\nonumber\\
 &= \frac{{\cal A}}{2\pi R \ell_1 \sqrt{2k} }\sum_n \int \frac{d\tau_2'}{(\tau_2')^{3/2}} 
 e^{-\frac{\left(R(\tau_2-\tau_2')+\pi\ell_1(Q_n-\bar Q_n)\tau_2'\right)^2}{8 \pi k \ell_1^2}+ 2\pi i k_n \tau_1-2\pi\tau_2' \varepsilon^0_n + 2\pi \tau_2' k_n \ell_1\frac{Q_n+\bar Q_n}{R}}\\
& = \sum_n e^{ 2\pi i k_n \tau_1 - 2\pi \tau_2 \varepsilon_n}\nonumber
\end{align}
with deformed energies 
\be
\varepsilon_n = \frac{R}{8\pi^2 k \ell_1^2}\left(R-\pi\ell_1 (Q_n-\bar Q_n) - \sqrt{(R-\pi\ell_1 (Q_n-\bar Q_n))^2 -16 \pi^2 k \ell_1^2(\varepsilon^0_n-\pi\ell_1 P_n (Q_n+\bar Q_n)) } 
\right)
\ee
with $P_n= k_n/R$.

This is the same as \eqref{eqn:jtbar-tjbar-energies} with the identification $\ell_{1} = \ell_{2}$, showcasing the somewhat surpising fact that deforming along one flow with the vector current is the same as deforming along two flows with the two chiral components.

\end{document}